\documentclass[11pt]{article}
\usepackage[margin=1in]{geometry}
\usepackage{mathptmx,bm}
\usepackage{graphicx,color, cite}
\usepackage{xspace,amsmath,amsfonts,amsthm,float,verbatim,enumitem,caption,xparse,cite}
\usepackage{caption}
\usepackage{hyperref}
\usepackage[capitalise]{cleveref}

\usepackage{tikz}
\usetikzlibrary{arrows,matrix,backgrounds}

\setlist[itemize]{leftmargin=*}

\title{
Semi-Streaming Algorithms for Annotated Graph Streams
}
\date{}

\author{
Justin Thaler%
\thanks{Yahoo Labs. The majority of this work was performed while the author was at the Simons Institute for the Theory of Computing, UC Berkeley. Supported by a Research Fellowship from the Simons Institute for the Theory of Computing.}
}

\newcommand{\fn}{F}

\newcommand{\field}{\mathbb{F}}

\newcommand{\pq}{{\sc PointQuery}\xspace}
\newcommand{\mysubset}{{\sc Subset}\xspace}
\newcommand{\odd}{{\sc odd}\xspace}

\newcommand{\triangles}{\textsc{Triangles}\xspace}
\newcommand{\mm}{\textsc{MaxMatching}\xspace}
\newcommand{\connectivity}{\textsc{Connectivity}\xspace}
\newcommand{\connectivitycc}{\textsc{Connectivity}$^{\text{cc}}$\xspace}
\newcommand{\disconnectivity}{\textsc{Disconnectivity}\xspace}
\newcommand{\disconnectivitycc}{\textsc{Disconnectivity}$^{\text{cc}}$\xspace}
\newcommand{\bipartiteness}{\textsc{Bipartiteness}\xspace}

\newcommand{\bipartitenesscc}{\textsc{Bipartiteness}$^{\text{cc}}$\xspace}

\DeclareMathOperator{\poly}{poly}

\DeclareMathOperator{\polylog}{polylog}

\DeclareMathOperator{\err}{err}
\DeclareMathOperator{\hcost}{hc}
\DeclareMathOperator{\vcost}{vc}

\DeclareMathAlphabet{\mathpzc}{OT1}{pzc}{m}{it}
\DeclareMathAlphabet{\mathcal}{OMS}{cmsy}{m}{n}

\newcommand{\ccfont}{\mathbf}
\DeclareDocumentCommand{\D}{m g}{%
  \IfNoValueTF{#2}%
    {\ccfont{\operatorname{\ccfont{D}}^{[#1]}}}%
    {\operatorname{D}^{[#1]}(#2)}%
}
\DeclareDocumentCommand{\R}{m g}{%
  \IfNoValueTF{#2}%
    {\ccfont{\operatorname{\ccfont{R}}^{[#1]}}}%
    {\operatorname{R}^{[#1]}(#2)}%
}
\DeclareDocumentCommand{\AM}{g}{%
  \IfNoValueTF{#1}%
    {\operatorname{\ccfont{AM}}}%
    {\operatorname{AM}(#1)}%
}
\DeclareDocumentCommand{\MA}{g}{%
  \IfNoValueTF{#1}%
    {\operatorname{\ccfont{MA}}}%
    {\operatorname{MA}(#1)}%
}
\DeclareDocumentCommand{\rMA}{m g}{%      <-- round-limited MA (Alice/Bob interaction)
  \IfNoValueTF{#2}%
    {\ccfont{\operatorname{\ccfont{MA}}^{[#1]}}}%
    {\operatorname{MA}^{[#1]}(#2)}%
}
\DeclareDocumentCommand{\OMA}{m g}{%      <-- online MA with Bob/Merlin interaction
  \IfNoValueTF{#2}%
    {\ccfont{\operatorname{\ccfont{OMA}}^{[#1]}}}%
    {\operatorname{OMA}^{[#1]}(#2)}%
}
\DeclareDocumentCommand{\OIP}{m g}{%
  \IfNoValueTF{#2}%
    {\ccfont{\operatorname{\ccfont{OIP}}^{[#1]}}}%
    {\operatorname{OIP}^{[#1]}(#2)}%
}
\DeclareDocumentCommand{\OIPS}{m g}{%
  \IfNoValueTF{#2}%
    {\ccfont{\operatorname{\ccfont{OIP}}^{[#1]}_{\bm{+}}}}%
    {\operatorname{OIP}^{[#1]}_{+}(#2)}%
}
\DeclareDocumentCommand{\OIPempty}{m g}{%
  \IfNoValueTF{#2}%
    {\ccfont{\operatorname{\ccfont{OIP}}^{#1}}}%
    {\operatorname{OIP}^{#1}#2}%
}
\DeclareDocumentCommand{\OMAempty}{m g}{%
  \IfNoValueTF{#2}%
    {\ccfont{\operatorname{\ccfont{OMA}}^{#1}}}%
    {\operatorname{OMA}^{#1}#2}%
}

\newcommand{\NC}{\operatorname{\ccfont{NC}}}

\newcommand{\out}[2]{\operatorname{out}^{#1}(#2)}

\newcommand{\ascheme}{scheme\xspace}

\newcommand{\aSchemes}{Schemes\xspace}

\newcommand{\eat}[1]{}
           
\renewcommand{\b}{\{0,1\}}

\newcommand{\help}{\mathfrak{h}}
\newcommand{\rhelp}{{\help}}

\newcommand{\alg}{\mathcal{A}}

\newcommand{\cP}{\mathcal{P}}

\newcommand{\cQ}{\mathcal{Q}}

\newcommand{\fail}{\bot}

\newcommand{\idx}{\textsc{index}\xspace}

\newcommand{\universe}{{\mathcal U}}

\newcommand{\etal}{{et al.}\xspace}
\newcommand{\mnote}[1]{}

\newcommand{\ignore}[1]{}
\newcommand{\provisionallyremove}[1]{}

\newcommand{\bx}{\mathbf{x}}

\newcommand{\anncost}{{c_a}}
\newcommand{\vercost}{{c_v}}
 
\newcommand{\nodes}{{n}}
\newcommand{\length}{{m}}
\newcommand{\stream}{{\sigma}}

\newtheorem{theorem}{Theorem}[section]

\newtheorem{corollary}[theorem]{Corollary}

\theoremstyle{definition}

\makeatletter
\newcommand\squeezepar{\@startsection{paragraph}{4}{\z@}{1.5ex \@plus1ex \@minus.2ex}{-1em}{\normalfont\normalsize\bfseries}}
\makeatother
\renewcommand{\paragraph}[1]{\squeezepar{{#1}}}

\begin{document}
\maketitle
\thispagestyle{empty}
\begin{abstract}

Considerable effort has been devoted to the development of
streaming algorithms for analyzing massive graphs.
Unfortunately, many results have been negative, establishing
that a wide variety of problems require  $\Omega(n^2)$ space
to solve. One of the few
bright spots has been the development of 
\emph{semi-streaming} algorithms for a handful of graph
problems --- these algorithms use space $O(n \cdot \polylog(n))$.

In the annotated data streaming model of Chakrabarti et al. \cite{icalp},
a computationally limited client wants to compute some property
of a massive input, but 
lacks the resources to store even a small fraction of the input, and hence
cannot perform the desired computation locally.
 The client therefore
accesses a powerful but untrusted
service provider, who not only performs the requested computation, 
but also \emph{proves} that
the answer is correct. 

We consider the notion
of \emph{semi-streaming algorithms for annotated graph streams} (semi-streaming annotation schemes for short). These are protocols in which both the client's space usage and the 
length of the proof are  $O(n \cdot \polylog(n))$.
We give evidence that semi-streaming annotation schemes represent a substantially
more robust solution concept than does the standard semi-streaming model.
On the positive side, we give semi-streaming annotation schemes for 
two dynamic graph problems that are intractable in the standard model: (exactly) counting triangles, and (exactly)
computing maximum matchings. The former scheme answers a question of Cormode \cite{openproblem}. On the negative side, we identify for the first time two natural graph problems (connectivity and bipartiteness in a certain edge update model) that can be solved in the standard 
semi-streaming model, but cannot be solved by annotation schemes
of ``sub-semi-streaming'' cost. That is, these problems are just as hard in the annotations model as they are in the standard model.
\end{abstract}
\newpage

\addtocounter{page}{-1}
%%%% \linenumbers %%%%
\section{Introduction}
The rise of cloud computing has motivated substantial interest in protocols
for \emph{verifiable data stream computation}. These protocols allow a computationally weak client (or \emph{verifier}), 
who lacks the resources to locally store a massive input,
to outsource the storage and processing of that input to a powerful but untrusted service provider
(or \emph{prover}). Such protocols provide a guarantee that the answer returned by the prover is correct, while allowing
the verifier to make only a single streaming pass over the input. %In practice,
%the verifier's streaming pass over the input can occur while the verifier is uploading the data to the cloud; 

Several recent works have introduced closely related models capturing the above scenario \cite{gur, icalp, ccc, prakash, soda, prakashnew, esa, vldb, itcs}.
Collectively, these works have begun to reveal a rich theory, leveraging algebraic
techniques developed in the classical theory of interactive proofs \cite{lfkn,
shamir92, babai, gmr} to obtain efficient verification protocols for a variety
of problems that require linear space in the standard streaming
model (sans prover). 

The primary point of difference among the various models of verifiable stream computation
that have been proposed is the amount of interaction
that is permitted between the verifier and prover.
The \emph{annotated data streaming} model of Chakrabarti \etal
\cite{icalp} (subsequently studied in \cite{esa, prakash, soda, itcs}) is non-interactive, requiring
the correctness proof to consist of just a single message from the prover to
the verifier, while other models, such as the \emph{Arthur--Merlin streaming protocols} of
Gur and Raz \cite{gur, soda} and the \emph{streaming interactive proofs} of Cormode et al. \cite{vldb, ccc}
permit the prover and verifier to exchange two or more messages. 
Our focus in this paper is on the annotated data streaming model of Chakrabarti \etal\ --- owing
to their non-interactive nature, protocols in this model possess a number of desirable properties not shared by their
interactive counterparts, such as reusability (see Section \ref{sec:goodprops} for details).
We are specifically concerned with protocols for problems on graph streams, described below.

\paragraph{Graph Streams.}
The ubiquity of massive relational data sets (derived, e.g. from the Internet and social networks) has 
led to detailed studies of data streaming algorithms for analyzing graphs. 
In this setting, the data stream consists of a sequence of edges, defining a graph $G$ on $n$ nodes, and the goal is to
compute various properties of $G$ (is $G$ connected? How many triangles does $G$ contain?). 
Unfortunately, many results on graph streaming have been negative: essentially any graph problem
of the slightest practical interest requires $\Omega(n)$ space to solve in the standard streaming model,
and many require $\Omega(n^2)$ space even to approximate. Due to their
prohibitive cost in the standard streaming model, many basic graph problems are ripe for outsourcing.

One of the few success stories in the study of graph streams has been the identification of the \emph{semi-streaming}
model as something of a ``sweet spot'' for streaming algorithms\cite{muthu, andrew}. 
The semi-streaming model is characterized by an $O(n \cdot \polylog n)$ 
space restriction, i.e., space proportional to the number of nodes rather than the number of edges. For dense
graphs this represents considerably less space than that required to store the entire graph. 
It has long been known that problems like connectivity and bipartiteness possess semi-streaming algorithms 
when the stream consists only of edge insertions, with no deletions. Recently, 
semi-streaming algorithms have been provided for these and other problems even for
dynamic graph streams, which contain edge deletions as well as insertions \cite{graphsketch1, graphsketch2, graphsketch3}. 
We direct the interested reader toward the recent survey of McGregor \cite{newsurvey} on graph stream algorithms.

In this work, we consider the notion of \emph{semi-streaming annotation schemes} for graph problems.
Here, the term ``scheme'' refers to a protocol in the annotated data streaming model. A scheme's total cost
is defined to be the sum of the verifier's space usage (referred to as the \emph{space cost} of the scheme) and the length of the proof (referred
to as the scheme's \emph{help cost}).
A scheme is said to be semi-streaming if its total cost is
$O(n \cdot \polylog n)$. 

We give evidence that semi-streaming annotation schemes represent a substantially more robust 
solution concept (i.e., a ``sweeter spot'') for graph problems than does the standard semi-streaming model. First, we give novel semi-streaming annotation schemes
for two challenging dynamic graph problems, counting triangles and maximum matching,
that require $\Omega(n^2)$ space in the standard streaming model. The total cost of these schemes is provably optimal up to a logarithmic factor.

Second, we show that two canonical problems that do possess semi-streaming algorithms
in the standard streaming model (connectivity and bipartiteness in a certain edge update model) are 
\emph{just as hard} in the annotations model. Formally, we show that any scheme for these problems
with space cost $O(n^{1-\delta})$ requires a proof of length $\Omega(n^{1+\delta})$ for any $\delta > 0$. 
Thus, for these problems, giving a streaming algorithm access to an untrusted prover
does not allow for a significant reduction in cost relative to what is achievable without a prover. This gives further evidence for the robustness of semi-streaming annotation schemes as a solution concept: while several fundamental problems that cannot be solved by standard semi-streaming algorithms can be solved by semi-streaming annotation schemes, there are
``easy'' problems (i.e., problems that do have semi-streaming solutions in the standard model) that cannot be solved by schemes of ``sub-semi-streaming'' cost.

\subsection{Summary of Contributions and Techniques}
%We now give a more detailed overview of our results, and a high-level sketch of the techniques
%we use to obtain them. 
Throughout this informal overview, $n$ will denote the number of nodes in
the graph defined by the data stream, and $m$ the number of edges. 
%A scheme with space cost upper bounded by $O(\vercost)$ and help cost
%upper bounded by $O(\anncost)$ is called a $(\anncost, \vercost)$ scheme.
To avoid boundary cases in the statement of our lower bounds,
we assume that the help cost of any scheme is always at least 1 bit.

\subsubsection{New Semi-Streaming Annotation Schemes}
Prior work has given semi-streaming annotation schemes for two specific graph problems that
require $\Omega(n^2)$ space in the standard semi-streaming model: bipartite perfect matching \cite{icalp}, and shortest $s$-$t$ path in graphs of polylogarithmic diameter \cite{esa}.
As discussed above, we give semi-streaming annotation schemes for two more challenging graph problems: maximum matching (\mm) and counting triangles (\triangles). Both schemes apply to dynamic graph streams.

\paragraph{Scheme for Counting Triangles.}

\begin{theorem}[Informal Version of Theorem \ref{thm:triangles}] \label{thm:triangles-overview}
  There is a scheme for \triangles with total cost $O(\nodes \log \nodes)$. Every scheme requires
  the product of the space and help costs to be $\Omega(n^2)$, and hence requires total cost $\Omega(n)$.
\end{theorem}

\begin{table}
\centering
\begin{tabular}{|c|c|c|}
\hline
Reference & \triangles\ Scheme Costs (help cost, space cost) & Total Cost Achieved\\
\hline
\cite{icalp} & $(n^2 \log n, \log n)$ & $O(n^2 \log n)$\\
\hline
\cite{icalp} & $(x \log n, y \log n)$ for any $x \cdot y \geq n^{3}$ & $O(n^{3/2}\log n)$\\
\hline
Theorem \ref{thm:triangles-overview} & $(n \log n, n \log n)$ & $O(n \log n)$\\
\hline
\end{tabular}
\caption{Comparison of our new scheme for \triangles\ to prior work.}
\label{tab:triangles}
\end{table}

Theorem \ref{thm:triangles-overview}
affirmatively answers a question of Cormode~\cite{openproblem},
resolves the Merlin-Arthur communication complexity of the problem up to a logarithmic
factor, and improves over the best previous upper bound of $O(n^{3/2} \log n)$, due to Chakrabarti et al. \cite{icalp} (see Table \ref{tab:triangles} for a detailed comparison to prior work).

As is the case for essentially all non-trivial protocols for verifiable stream computation, the scheme of Theorem \ref{thm:triangles-overview} uses algebraic techniques related to
the famous \emph{sum-check protocol} of Lund \etal \cite{lfkn} from the classical theory of interactive proofs. Yet, our scheme deviates in a significant way from all
earlier annotated data stream and interactive proof protocols \cite{gkr, ccc, icalp, shamir92}.  Roughly speaking, in previous protocols, the
verifier's updates to her memory state were commutative, in the sense that
reordering the stream tokens would not change the final state reached by the
verifier. However, our new verifier is inherently non-commutative:  her
update to her state at time $i$ depends on her actual state at time $i$,
and reordering the stream tokens can change the final state reached by the verifier.
See Section
\ref{sec:trianglescomparison} for further discussion of this point.  

Like our scheme for \mm\ below, our scheme for \triangles\ does not achieve smooth tradeoffs between space and help costs:
we do not know how to reduce the space usage to $o(n \log n)$ without blowing
the help cost up to $\Omega(n^2)$, or vice versa.  This is in contrast
to prior work on annotated data streams \cite{icalp, esa, soda, gur}, which
typically achieved any combination of space and help costs subject to the
product of these two costs being above some threshold.  We conjecture that
achieving such smooth tradeoffs for either problem is impossible.

\paragraph{Scheme for Maximum Matching.}

\begin{theorem} \label{thm:mmoverview}[Informal Version of Theorem \ref{thm:mm}] 
There is a scheme for \mm\ with total cost $O(n \log n)$. 
  Every scheme for \mm\ requires
  the product of the space and help costs to be $\Omega(n^2)$, and hence requires total cost $\Omega(n)$.
\end{theorem}

\begin{table}
\centering
\begin{tabular}{|c|c|c|}
\hline
Reference & \mm\ Scheme Costs (help cost, space cost) & Total Cost Achieved\\
\hline
\cite{esa} & $(m \log n, \log n)$ & $O(m \log n)$\\
\hline
Theorem \ref{thm:maxmatching-overview} & $(n \log n, n \log n)$ & $O(n \log n)$\\
\hline
\end{tabular}
\caption{Comparison of our new scheme for \mm\ to prior work.}
\label{tab:mm}
\end{table}

Our scheme combines the Tutte-Berge formula with algebraic techniques 
to allow the prover to establish matching upper and lower bounds on the size of a maximum matching
in the input graph.
Schemes for maximum matching had previously been studied by Cormode \etal \cite{esa},
but this prior work only gave schemes with help cost proportional to the number of edges, which
is $\Omega(n^2)$ in dense graphs (like us, Cormode \etal exploited the Tutte-Berge formula, but  
did not  do so in a way that achieved help cost sublinear in the input size). Prior work had also given a scheme achieving optimal
tradeoffs between help and space costs for \emph{bipartite perfect matching} \cite[Theorem 7.5]{icalp} -- 
our scheme for \mm\ can be seen as a broad generalization of \cite[Theorem 7.5]{icalp}.

\subsubsection{New Lower Bounds} 
\label{sec:lboverview}
On the other hand, we identify, for the first time, natural graph 
problems that possess standard semi-streaming algorithms, 
but in a formal sense are just as hard in the annotations model as they are in the standard streaming model. 
The problems that we consider are connectivity and bipartiteness in a certain edge update model that we call the XOR update model.
In this update model,
the stream $\langle e_1, \dots, e_{\length} \rangle$ is a sequence of edges from $[n] \times [n]$, which
define a graph $G = (V, E)$ via: $e \in E \Longleftrightarrow |\{i: e_i = e\}| = 1 \mod 2$. Intuitively,
each stream update $e_i$ is interpreted as changing the status of edge $e_i$: if it is currently in the graph, then the update causes $e_i$ to be deleted; otherwise $e_i$ is inserted. 
Our lower bound holds for schemes for connectivity and bipartiteness in 
the XOR update model, even under the promise that $e_1, \dots, e_{\length - n}$ are all unique (hence, all but the last $n$ stream updates
correspond to edge insertions), and the last $n$ updates are all incident to a single node.

\eat{
\begin{table}
\centering
\begin{tabular}{|c|c|c|c|} 
\hline
Reference & Upper or Lower Bound & Scheme Costs (help cost, space cost) & Total Cost Achieved\\
\hline
\cite{graphsketch1} & Upper Bound & $(1, n \polylog(n))$ & $O(n \polylog(n))$\\
\hline
\cite{icalp} & Upper Bound & $(m \log n, \log n)$ & $O(m \log n)$\\
\hline
\cite{icalp} & Upper Bound & $(x \log n, y \log n)$ for any $x \cdot y \geq n^{2}$ & $O(n \log n)$\\
\hline
\cite{icalp} & Lower Bound & $\anncost \cdot \vercost = \Omega(n)$ & $\Omega(n^{1/2})$\\
\hline
Theorem \ref{thm:connectivity} & Lower Bound & $(n + \anncost) \cdot \vercost = \Omega(n^2)$ & $\Omega(n)$\\
\hline
\end{tabular}
\label{tab:connectivity}
\caption{Comparison of our new lower bound for \connectivity\ and \bipartiteness\ in the XOR
update model, with prior work. The first three entries in the table refer to upper 
bounds, while the last two entries refer to lower bounds on the space and help costs of any 
scheme. All bounds from prior work were originally presented for the turnstile update model,
but can be modified to apply to the XOR update model. 
}
\end{table}
}

\begin{theorem}[Informal Version of Corollary \ref{cor:lb}] 
\label{thm:connectivity}
Consider any scheme for \connectivity\ or \bipartiteness\ in the XOR update model with help cost $\anncost$
and and space cost $\vercost$. Then $(\anncost+n) \cdot \vercost  \geq n^2$, 
even under the promise that the first $\length - n$ stream updates are all unique, and the last $n$ stream updates are all incident to a single node. In particular, the total cost of any annotation scheme for these problems is $\Omega(n)$. 
\end{theorem}

Both connectivity and bipartiteness in the XOR update model possess standard semi-streaming algorithms \cite{graphsketch1}.\footnote{The algorithms of \cite{graphsketch1} are
described in the turnstile update model, in which each stream update explicitly specifies whether to insert or delete a (copy of) an edge. However, these algorithms are easily modified 
to apply to the XOR update model as well. In brief, these algorithms have
$L_0$-sampling algorithms at their core. Existing $L_0$-samplers are in turn built on top of sparse recovery algorithms (see, e.g., \cite{grahamsampling}), and many sparse recovery algorithms can be
implemented in the XOR update model directly (see, e.g., \cite{iblt}).}
Hence, Theorem \ref{thm:connectivity}
implies that the total cost of any annotation scheme is at most a polylogarithmic factor smaller than the problems' space complexity in the standard streaming model.
Like all prior work establishing lower bounds on the cost of protocols for verifiable stream computation, our lower bounds 
are established using notions of \emph{Merlin-Arthur communication complexity} \cite{icalp, esa, prakash, ccc, gur}.

Prior to this work, only one other problem was known to be as hard (up to logarithmic factors) in
the annotations model as in the standard streaming model \cite[Corollary 3.3]{soda}.
The problem considered in \cite[Corollary 3.3]{soda} was an ``exponentially sparse'' variant of the 
classic \idx\ problem, in which the data stream consists of a vector $x \in \{0, 1\}^n$ 
promised to have Hamming weight $O(\log n)$, followed by an index $i \in [n]$, and the
goal is to output the value $x_i$. 
Connectivity and bipartiteness are arguably more natural problems, and
are qualitatively different, as we now explain.

On an informal level, the reason the exponentially sparse \idx\ problem is hard 
in the annotations model is that any ``useful''
annotation must at least specify a single index into the vector $x$, which requires $\log n$ bits of annotation.
And since $x$ is exponentially sparse, $\log n$ is actually equal (up to a constant factor) to the space complexity of a standard streaming algorithm for the problem. In our view, \bipartiteness\ and \connectivity\ are hard in the annotations
for a different reason -- roughly speaking, any useful annotation for these problems must at least specify, for each node $u$, the side of the bipartition or the connected component in which $u$ resides.

%All other known lower bounds on the total cost of annotation schemes have left at least a quadratic gap between the lower bound, and the space complexity in the standard streaming model of the problem of interest. 

\medskip \noindent \textbf{Overview of the Proof.} Our proof of Theorem \ref{thm:connectivity} works by specifying a reduction from the
\idx\ problem on inputs of length $n^2$, for which a lower bound of $\Omega(n^2)$ on the product
of the help and space costs of any annotation scheme was established in \cite{icalp},
to \connectivity\ and \bipartiteness\ on graphs with $n$ nodes and $\Theta(n^2)$ edges.

Notice that in the standard (sans prover) streaming model, the \idx\ problem on $n^2$ variables is strictly harder
than connectivity and bipartiteness problems on graphs with $n$ nodes, as the former requires $\Omega(n^2)$ space, while the latter two problems require only $O(n \cdot \polylog(n))$ space.
Yet Theorem \ref{thm:connectivity} establishes that in the annotations model,
all three problems are of essentially equivalent difficulty (in particular, schemes of total cost $\tilde{O}(n)$ are necessary and sufficient to solve all three problems). 
To establish such a result, it is necessary to use a reduction 
that is specifically tailored to the annotations model, in the sense that the reduction
must not apply in the standard streaming model (since \idx\ and \connectivity\
are not of equivalent difficulty in the standard setting). Namely, in our reduction
from \idx\ to connectivity, the prover
\emph{helps the verifier} transform an instance of the \idx\ problem into a \connectivity\ instance. 
This ``help'' consists of $\Theta(n)$ bits, and this is why our lower bound is of the form
$(\anncost + n) \cdot \vercost \geq n^2$. This is in contrast to prior lower bounds,
which, with the exception of \cite[Corollary 3.3]{soda}, were all of the form $\anncost \cdot \vercost = \Omega(C)$ for some quantity $C$. 
 
%We conjecture that the lower bound of Theorem \ref{thm:connectivity} holds even in the strict turnstile update model, as opposed to the XOR update model. 

\subsection{Other Related Work}
\eat{As described earlier, the annotated data steaming model was proposed by Chakrabarti et al. \cite{icalp}, who gave 
essentially optimal or nearly optimal schemes for a wide variety of problems, including frequency moments and several graph problems.
Subsequent work by Cormode et al. \cite{esa} presented schemes for several more problems, with a particular focus on graph problems.
The schemes we give in this work directly improve or generalize schemes from both of these earlier papers.
Several subsequent papers studied related models of 
verifiable stream computation \cite{itcs, ccc, vldb, prakash, prakashnew, gur, soda}. Refinements and implementations
\cite{allspice,itcs,thaler} have demonstrated genuine practicality for many of the protocols developed in this line of work.}

As discussed above, several recent papers \cite{itcs, ccc, vldb, prakash, prakashnew, gur, soda, icalp, esa} have all studied
annotated data streams and closely related models for verifiable stream computation. 
Refinements and implementations
\cite{allspice,itcs,thaler} have demonstrated genuine practicality for many of the protocols developed in this line of work.

Protocols for verifiable stream computation have also been studied 
in the cryptography community \cite{chung, eurocryptdie, schroder}.
These works only require security against cheating provers that run in polynomial time, as compared
to the setting we consider, where security holds even against computationally unbounded provers.
In exchange for the weaker security guarantee, these protocols may achieve properties
that are unattainable in our information-theoretic setting. 
For example, some of these protocols achieve a stronger form of reusability than we do (see Section \ref{sec:goodprops} for our definition of reusability) --- they remain secure
for many uses even if the prover learns all of the verifier's accept/reject decisions. The work of Chung et al. \cite{chung}
uses fully homomorphic encryption (FHE), which remains far from practical at this time. Schr\"{o}der and Schr\"{o}der \cite{schroder},
and Papamanthou et al. \cite{eurocryptdie} avoid the use of FHE, but handle only much simpler queries (such as point queries and range search)
than the graph problems we consider here.

\vspace{-4mm}
\section{Models of Streaming Computation}
\vspace{-2mm}
Our presentation of data streaming models closely follows Chakrabarti et al. \cite{soda}.
Recall that a (standard) data stream algorithm computes a function $f$ of an input sequence
$\bx \in\universe^\length$, where $\length$ is the number of stream updates, and $\universe$ is some data universe.
The algorithm has only sequential access to $\bx$, uses a limited amount of space, and has
access to a random string. The function $f$ may or may not be Boolean.
%When it is not, we may permit an approximate answer: we accept any answer in the set $C(f(\bx))$, for some function $C$. 

%The annotated data streaming model of Chakrabarti et al. \cite{icalp} 
%is defined as follows.
An annotated data stream algorithm, or a {\em \ascheme}, is a pair $\alg =
(\help,V)$, consisting of a help function $\help:\universe^\length \times \b^*
\to \b^*$ used by a {\em prover} and a data stream algorithm
run by a {\em verifier}, $V$. The prover provides $\help(\bx)$ as
annotation to be read by the verifier.  We think of $\help$ as being decomposed into
$(\help_1,\ldots,\help_{\length})$, where the function
$\help_i:\universe^{\length}\to\b^*$ specifies the annotation supplied
after the arrival of the $i$th token $x_i$. That is, $\help$ acts on
$\bx$ to create an {\em annotated stream} $\bx^\rhelp$ defined as follows:
\vspace{-2mm}
\[
  \bx^{\rhelp} :=
  (x_1,\, \help_1(\bx),\,
    x_2,\, \help_2(\bx),\, \ldots,\,
    x_{\length},\, \help_{\length}(\bx)) \, .
\]

Note that this is a stream over $\universe \cup \b$, of length $\length +
\sum_i |\help_i(\bx)|$.  The streaming verifier, who has access to a (private) random string $r$, then
processes this annotated stream, eventually giving an output
$\out{V}{\bx^{\rhelp},r}$.

% For ease of exposition, as well as consistency with the literature on interactive proofs,
% we will regularly refer to the help function $\help$ as a \emph{prover}, which we denote by $P$.
% Likewise, we will refer to the data stream algorithm $V$ as the \emph{verifier}.

\paragraph{Online \aSchemes.}
We say a scheme is \emph{online} if each function $\help_i$
depends only on $(x_1, \ldots, x_i)$.
The \ascheme\ $\alg = (\help,V)$ is said to be $\delta_s$-sound and
$\delta_c$-complete for the function $\fn$ if the following conditions hold:
\begin{enumerate}
\vspace{-1mm}
  \item For all $\bx\in\universe^\length$, we have $\Pr_{r}[\out{V}{\bx^{\rhelp},r} \neq \fn(\bx)] \le \delta_c$.
  \vspace{-1mm}
  \item For all $\bx\in\universe^\length$, $\help'=(\help_1', \help_2',
  \ldots, \help_\length') \in (\b^*)^\length $, we have
    $\Pr_{r}[\out{V}{\bx^{\help'},r} \not\in \{\fn(\bx)\} \cup\{\fail\}] \le \delta_s$.
    \vspace{-1mm}
\end{enumerate}
\vspace{-1mm}
 An output of ``$\fail$''
indicates that the verifier rejects the prover's claims in trying to convince the verifier to output a
particular value for $\fn(\bx)$.  
We define $\err(\alg)$ to be the minimum value of $\max\{\delta_s, \delta_c\}$
such that the above conditions are satisfied. We define the {\em annotation
length} $\hcost(\alg) = \max_{\bx} \sum_i |\help_{i}(\bx)|$, the
total size of the prover's communications, and the {\em verification space cost}
$\vcost(\alg)$ to be the space used by the verifier.  We say that $\alg$ is
an online ($\anncost, \vercost)$ scheme if $\hcost(\alg) = O(\anncost)$, $\vcost(\alg) =
O(\vercost)$, and $\err(\alg) \le \frac13$ (the constant 1/3 is arbitrary and chosen
by convention).

\vspace{-1mm}
\paragraph{Prescient \aSchemes.}
Chakrabarti et al. \cite{icalp} also define the notion of a \emph{prescient} scheme, which is
the same as an online scheme, except the annotation at any time $i$ is allowed to
depend on data which the verifier has not seen yet.  Prescient schemes have the undesirable property
that the prover may need to ``see into the future'' to convince the verifier to produce the correct output.
%All of the schemes we present in this paper are online. 
Note that
even though our \triangles\ and \mm\ protocols are online, they
are optimal up to logarithmic factors \emph{even among prescient schemes} (see Theorems \ref{thm:triangles} and
\ref{thm:mm} for details). 
\vspace{-1mm}
\paragraph{Additional Properties of Our Schemes.}
\label{sec:goodprops}
While the annotated data streams model allows the prover to interleave the annotation with the stream,
in all of the schemes we present in this paper, all of the annotation comes at the end of the stream. This property
avoids any need for fine-grained coordination between the annotation and the stream, and permits the prover
to send the annotation as a single email attachment, or post it to a website for the verifier to retrieve at her convenience. 
We clarify that the \emph{lower bounds} for \connectivity\ and \bipartiteness\ that we establish in Section \ref{sec:lb} 
apply to any online scheme, even those which interleave the annotation with the stream.

The schemes we present in this work permit a natural form of \emph{reusability}: 
if the verifier wants to compute a function $f$ on a given dataset $\bx$, 
the verifier can receive $f(\bx)$  (with a correctness proof), and check the validity of the proof using 
a ``secret state'' that she computed while observing the stream $\bx$.
Further updates to the stream $\bx$ can then occur, yielding a (longer) stream $\bx'$, and 
the verifier can update her secret in a streaming fashion.
The verifier may then receive the answer $f(\bx')$ (with a correctness proof) on the updated dataset, and check
its correctness using the updated secret state. The probability the verifier gets fooled into outputting an incorrect answer 
on even a single query grows only linearly with the number of times the prover sends the verifier an answer.
This kind of reusability is not possible with many interactive solutions \cite{vldb, ccc, prakashnew}, which typically require
the verifier to reveal information about $r$ to the prover over the course of the protocol.

%\vspace{-5mm}
\section{Upper Bounds}
%\vspace{-2mm}
\paragraph{Graph Streams in the Strict Turnstile Model.} 
The annotation schemes of this section apply to graph streams in the strict turnstile update model.
In this model, a data stream $\stream$ consists of a sequence of undirected edges, accompanied by (signed) multiplicities: $\langle (e_1, \Delta_1), \dots, (e_{\length}, \Delta_{\length}) \rangle$. Each edge $e_i \in [\nodes] \times [\nodes]$, and each $\Delta_i \in \mathbb{Z}$. An update $(e_i, \Delta_i)$ with $\Delta_i > 0$ is interpreted as an insertion of
$\Delta_i$ copies of edge $e_i$ into graph $G$. If $\Delta_i < 0$, the update is interpreted as a deletion of $\Delta_i$ copies of edge $e_i$.
It is assumed that at the end of the stream, no edge has been deleted more times than it has been inserted (all of our protocols
work even if this property does not hold at \emph{intermediate} time steps, as long as the property holds after the final stream update has been processed).\footnote{The reason that we do not consider the case where edges may have negative weights at the end of the stream is that it is unclear what is the most meaningful way to define problems like \triangles\ and \mm\ in this setting. See Footnotes 3 and 6.}
%For simplicity, we describe our algorithms assuming that, at the end of the stream, every edge has multiplicity at most $1$, but our protocols easily
%extend to handle multigraphs. 
%
When analyzing the time costs of our schemes, we assume that any addition
or multiplication in a finite field of size $\poly(n)$ takes one unit of time.

%\vspace{-3mm}
\subsection{A Semi-Streaming Scheme for Counting Triangles}
%\vspace{-2mm}
\label{sec:triangles}
In the \triangles problem, 
the goal is to determine the number of unordered triples of distinct vertices $(u, v, z)$ 
such that edges $(u,v)$, $(v, z)$, and $(z, u)$ all appear in $G$. 
More generally, if these edges appear with respective multiplicities $M_1$, $M_2$, and $M_3$, we view triple $(u,v,z)$ as contributing $M_1 \cdot M_2 \cdot M_3$ triangles to the total count.\footnote{The \triangles scheme of Theorem 3.1 gives meaningful results even if the $M_i$'s may be negative: a triangle with an odd (even) number of edges of negative multiplicity contributes a negative (positive) number to the total triangle count.}
Computing the number of triangles is a well-studied problem \cite{AlonYZ97} and there has been considerable interest in designing  algorithms in a  variety of models including the data stream model
\cite{Bar-YossefKS02,PavanTTW13}, MapReduce \cite{SuriV11}, and the quantum query model \cite{LeeMS13}. One motivation is the study of social networks where important statistics such as the clustering coefficient and transitivity coefficient are based on the number of triangles.
Understanding the complexity of counting triangles captures the
ability of a model to perform a non-trivial correlation within large
graphs. 
Chakrabarti \etal \cite{icalp} gave two annotated data streaming protocols
for this problem. The first protocol had help cost $O(n^2 \log \nodes)$,
and space cost $O(\log \nodes)$. The second protocol
achieved help cost  $O(x \log \nodes)$ and space cost $O(y \log \nodes)$ for  
any $x, y$ such that $x \cdot y \geq \nodes^3$. 
In particular, by setting $x=y=\nodes^{3/2}$, the second protocol of Chakrabarti \etal ensured that both help cost and space cost  equaled $O\left(\nodes^{3/2} \log \nodes\right)$.
Cormode \cite{openproblem} asked whether it is possible
to achieve an annotated data streaming protocol in which both the help cost and space cost are $\tilde{O}(\nodes)$. We answer this question in the affirmative. 
%\vspace{-1mm}
\begin{theorem}[Formal Statement of \Cref{thm:triangles-overview}] \label{thm:triangles}
 Assume there is an a priori upper bound $B \leq \poly(n)$ on the multiplicity of any edge in $G$. 
 There is an online scheme for \triangles\ with space and
  help costs $O(\nodes \log \nodes)$. Every scheme (online or prescient) requires
  the product of the space and help costs to be $\Omega(n^2)$, and hence total cost $\Omega(n)$, even for $B=1$, and even if $G$ is promised to have exactly 0 or 1 triangles.
\end{theorem}
%\vspace{-2mm}
\noindent \textbf{Discussion.} \label{sec:discuss} Before proving Theorem \ref{thm:triangles}, it is instructive
to consider the following
simple \emph{interactive} streaming verification protocol for \triangles. The approach is to apply the sum-check protocol
of Lund et al. \cite{lfkn} to a suitably defined multivariate polynomial $h$. The space and communication costs of this protocol are comparable to that of Theorem \ref{thm:triangles}; the advantage of Theorem \ref{thm:triangles} over this simple solution is that Theorem \ref{thm:triangles} gives a protocol that is \emph{non-interactive} (and comes with the associated reusability benefits described in Section \ref{sec:goodprops}). 

We direct the interested reader to \cite[Chapter 8]{arorabarak} for a detailed description of the sum-check protocol. For our purposes, its crucial properties are: (a) Given as input a $k$-variate polynomial $h$ defined over a field $\field$, and a subset $H \subseteq \field$, the sum-check protocol can compute the quantity $\sum_{x_1, \dots, x_k \in H} h(x_1, \dots, x_k)$. (b) The prover and verifier in the sum-check protocol exchange $2k-1$ messages. (c) The communication cost of the sum-check protocol is $O(k \cdot \deg(h))$ field elements, where $\deg(h)$ denotes the maximum degree of $h$ in any variable.
(d) In order to run her part of the sum-check protocol, the verifier chooses $k$ values $r_1, \dots, r_k$ at random from $\mathbb{F}$, and needs to evaluate
 $h(r_1, r_2, \dots, r_k)$.

Hence, to give an interactive protocol for \triangles, it suffices to identify a low-degree $k$-variate polynomial $h$, a finite field $\field$ of prime order, and 
a subset $H \subseteq \field$ such that the number of triangles in the input graph $G$ equals\footnote{Here, we are identifying natural numbers less than $|\field|$ with elements of $\field$ in the natural way.} $\sum_{x_1, \dots, x_k \in H} h(x_1, \dots, x_k)$. Moreover, the verifier must be able to
evaluate $h(r_1, \dots, r_k)$ in small space with a single streaming pass over the input. 
To this end, let $E(u, v) \colon [\nodes] \times [\nodes] \rightarrow \mathbb{Z}$ denote the function
that outputs the multiplicity of the edge $(u, v)$ in the graph $G$ defined by the input stream. 
%The number of triangles in $G$ equals $ \sum_{u, v, w \in [\nodes]} E(u, v) \cdot E(v, w) \cdot E(w, u).$
Let $\field$ denote a finite field of prime size $6 (B \cdot n)^3 \leq |\field| \leq 12 (B \cdot n)^3$, and let $\tilde{E}(X, Y)$ denote the unique
polynomial over $\field$ of degree 
at most $n$ 
in each variable $X, Y$ such that 
$\tilde{E}(u, v) = E(u, v)$ for all $(u, v) \in [\nodes] \times [\nodes]$. 
Let $h(X, Y)$ denote the
{bivariate} polynomial $h(X, Y) := \sum_{w \in [\nodes]} \tilde{E}(X, Y) \cdot \tilde{E}(Y, w) \cdot \tilde{E}(w, X),$ and note that
the number of triangles in $G$ is equal to $\sum_{u, v \in [\nodes]} h(u, v)$ as desired. 

By Property (b) above, applying the sum-check protocol to $h$ requires the prover and verifier to exchange three messages.
By Property (c) above, the communication cost is $O(n)$ field elements, since $h$ has degree at most $2n$ in $X$ and in $Y$. Finally, it is possible to show that the verifier can evaluate $h(r_1, r_2)$ for any points $r_1, r_2 \in \field$ with a single streaming pass over the input, using space $O(n \log |\mathbb{F}|)$ (we omit the details of this computation for brevity). 
In summary, we have given a three-message interactive streaming verification protocol for \triangles, with space and communication costs bounded by $\tilde{O}(n)$.\footnote{Similarly, a five-message protocol with communication cost $\tilde{O}(n)$ and space cost $\tilde{O}(\log n)$
can be obtained by applying the sum-check protocol to the {trivariate} polynomial
$h'(X, Y, Z) :=  \tilde{E}(X, Y) \cdot \tilde{E}(Y, Z) \cdot \tilde{E}(Z, X)$, to compute the quantity $\sum_{u, v, w \in [\nodes]} h'(u, v, w)$.}

%A \emph{three-message} streaming verification protocol for computing $\sum_{u, v \in [\nodes]} h(u, v)$ with communication cost bounded by $O(n \log n)$ 
%An \emph{interactive} streaming verification protocol for computing $\sum_{u, v \in [\nodes]} h(u, v)$ can be obtained by applying the sum-check protocol of Lund et al. \cite{lfkn} to $h(X, Y)$.
%We direct the interested reader to \cite[Chapter 8]{arorabarak} for a detailed description of the sum-check protocol. For our purposes, the crucial properties of the sum-check protocol are the following: (1) when applied to a $k$-variate polynomial, the sum-check protocol requires the prover and verifier to exchange $2k-1$ messages. Since $h(X, Y)$ is defined over $k=2$ variables, applying the sum-check protocol to $h$ requires the prover and verifier to exchange $3$ messages.\footnote{Similarly, a \emph{five-message} protocol with communication cost $\tilde{O}(n)$ and space cost $\tilde{O}(\log n)$
%can be obtained by applying the sum-check protocol to the {trivariate} polynomial
%$h'(X, Y, Z) :=  \tilde{E}(X, Y) \cdot \tilde{E}(Y, Z) \cdot \tilde{E}(Z, X)$, to compute the quantity $\sum_{u, v, w \in [\nodes]} h'(u, v, w)$.} 
% (2) In order to run her part of the sum-check protocol, the verifier randomly chooses two values $r_1, r_2$ at random from $\mathbb{F}$, and needs to evaluate the quantity 
% $h(r_1, r_2)$. It is possible to show that the verifier can do this in space $O(n \log |\mathbb{F}|)$ with a single streaming pass over the input.

In order to obtain the scheme of Theorem \ref{thm:triangles} below, we need to remove the interaction from the above 3-message protocol. To accomplish this, we identify a \emph{univariate} polynomial $g(Z)$ of degree $O(n)$ such that the number of triangles in $G$ equals
$\sum_{z \in [\nodes]} g(Z)$. Moreover, we show that the verifier can evaluate $g(r)$ for any point $r \in \field$  in space $O(n \log |\mathbb{F}|)$ with a single streaming pass over the input. It follows that applying the sum-check protocol to $g$ yields an scheme with costs claimed in Theorem \ref{thm:triangles}.
The polynomial $g$ that we identify is defined as a sum of $m$ constituent polynomials, one for each stream update. 

%\vspace{-2mm}
\begin{proof}[\textbf{Proof of Theorem \ref{thm:triangles}}]
The lower bound was proved in \cite[Theorem 7.1]{icalp}. Details of the upper bound follow.
Let $G_i$ denote the graph defined by the first $i$ stream updates $\langle (e_1, \Delta_1), \dots, (e_i, \Delta_i) \rangle$, and let $E_{i} \colon [\nodes] \times [\nodes] \rightarrow \mathbb{Z}$ denote the function
that outputs the multiplicity of the edge $(u, v)$ in graph $G_{i-1}$. 
On edge update $e_i = (u_i, v_i)$, notice that the number of triangles that $e_i$ \emph{completes}
in $E_i$
is precisely %\vspace{-2mm} $$\Delta_i \cdot \sum_{z \in [\nodes]} E_i(u_i, z) E_i(v_i, z).$$ Thus, 
the total number of triangles in the graph $G=G_{\length}$
is precisely $\sum_{i \leq \length} \Delta_i \sum_{z \in [\nodes]} E_i(u_i, z) E_i(v_i, z).$ % \end{equation}
%We use a novel variant of the sum-check protocol \cite{lfkn, aw} to compute this quantity.
%To this end, 
Let $\field$ denote
a field of prime order $6 (B \cdot n)^3 \leq |\field| \leq 12 (B \cdot \nodes)^3$, and
let $\tilde{E}_i(X, Y)$ denote the unique 
polynomial over $\field$ of degree 
at most $n$ 
in each variable $X, Y$ such that 
$\tilde{E}_i(u, v) = E_i(u, v)$ for all $(u, v) \in [\nodes] \times [\nodes]$. 
%
%denote the unique polynomial over $\field$ of degree at most $n$ 
%in each variable $X, Y$ such that 
%$\tilde{\delta}_{(u, v)}(x, y) = \delta(x, y)$ for all inputs $(x, y) \in [n] \times [n]$. 
%We refer to $\tilde{\delta}_{(i. j)}$ as a \emph{low-degree extension} of the function $\delta$. 
%
Then the number of triangles in $G$ equals
%\vspace{-2mm}
\begin{equation}
\label{eq:trianglespolys}\sum_{i \leq \length} \Delta_i \sum_{z \in [\nodes]} E_i(u_i, z) E_i(v_i, z) =
\sum_{i \leq \length} \Delta_i \sum_{z \in [\nodes]} \tilde{E}_i(u_i, z) \tilde{E}_i(v_i, z)
=  \sum_{z \in [\nodes]}  \sum_{i \leq \length} \Delta_i \cdot \tilde{E}_i(u_i, z) \tilde{E}_i(v_i, z).
 \end{equation}
 %%\vspace{-1mm}
In turn, the right hand side of Equation \eqref{eq:trianglespolys}
can be written as $\sum_{z \in [\nodes]} g(z)$, where
$g$ denotes the \emph{univariate} polynomial defined via: 
%\vspace{-2mm}
\begin{equation} \label{eq:g} g(Z)= \sum_{i \leq \length} \Delta_i \cdot \tilde{E}_i(u_i, Z) \tilde{E}_i(v_i, Z).\end{equation}
%%\vspace{-1mm}
Notice $g(Z)$ is a univariate polynomial of degree at most $2\nodes$. 
Our scheme proceeds as follows. 

\medskip
\noindent \textbf{Prover's computation.}
At the end of the stream, the prover sends a univariate polynomial $s(Z)$ of degree at most $2\nodes$,
where $s(Z)$ is
claimed to equal $g(Z)$. Notice that since $s(Z)$ has degree at most $2\nodes$,
$s(Z)$ can be specified by sending its values on all inputs in $\{0, \dots, 2\nodes\}$ --- this requires 
help cost  $O(n \log |\field|) = O(\nodes \log \nodes)$.

\medskip
\noindent \textbf{Verifier's computation.}
At the start of the stream, the verifier picks a random field element $r \in \field$,
and keeps the value of $r$ secret from the prover.
We will show below that the verifier can evaluate $g(r)$ with a single 
streaming pass over the input,
using space $O(\nodes \log \nodes)$.
The verifier checks whether $s(r) = g(r)$. If 
this check fails, the verifier halts and rejects. If the check passes, the verifier
outputs $\sum_{z \in [\nodes]} s(z)$ as the correct answer. 

We now explain how the verifier can evaluate $g(r)$ with a single streaming
pass over the input. The high-level idea is as follows. 
Equation \eqref{eq:g} expresses $g(r)$ as a sum of $\length$ terms,
where the $i$th term equals
$\Delta_i \cdot \tilde{E}_i(u_i, r) \tilde{E}_i(v_i, r)$.
For each $u \in [\nodes]$, we will show how the verifier can incrementally maintain 
the quantity $\tilde{E}_i(u, r)$ at all times $i$. The verifier
will maintain all $\nodes$ of these quantities, resulting in 
a total space cost of $O(\nodes \log|\field|) = O(\nodes \log \nodes)$.
With these quantities in hand, it is straightforward for 
the verifier to incrementally maintain the sum 
$\sum_{j \leq i} \Delta_j \cdot \tilde{E}_j(u_j, r) \tilde{E}_j(v_j, r)$ at all times $i$:
upon the $i$th stream update, the verifier simply
adds $\Delta_i \cdot \tilde{E}_i(u_i, r) \cdot \tilde{E}_i(v_i, r)$ to the running sum. 
 
To maintain the quantity 
$\tilde{E}_i(u, r)$,
we begin by writing the bivariate polynomial $\tilde{E}_i(X, Y)$ 
in a convenient form. 
Given a pair $(u, v) \in [\nodes] \times [\nodes]$,
%let $\delta_{(u, v)}(x, y) : [\nodes] \times [\nodes] \rightarrow \{0, 1\}$
%denote the function that evaluates to 1 on input $(u, v)$, and evaluates
%to 0 on all inputs $(x, y) \neq (u, v)$. 
let $\tilde{\delta}_{(u, v)}$ denote the following (Lagrange) polynomial:
$\tilde{\delta}_{(u, v)}(X, Y) = \left(\frac{\prod_{1 \leq u' \leq n: u' \neq u}(X-u')}{\prod_{1 \leq u' \leq n: u' \neq u}(u-u')}\right)\left(\frac{\prod_{1 \leq v' \leq n: v' \neq v}(Y-v')}{\prod_{1 \leq v' \leq n: v' \neq v}(v-v')}\right)$.
Notice that $\tilde{\delta}_{(u, v)}$ evaluates to 1 on input $(u, v)$, and evaluates
to 0 on all other inputs $(x, y) \in [\nodes] \times [\nodes]$.
Thus, we may write  
$\tilde{E}_i(X, Y) = \sum_{j \leq i} \tilde{\delta}_{(u_j, v_j)}(X, Y)$.
In particular, for each node $u \in [\nodes]$, $\tilde{E}_i(u, r) =
\tilde{E}_{i-1}(u, r) + \tilde{\delta}_{(u_i, v_i)}(u, r) +  \tilde{\delta}_{(v_i, u_i)}(u, r)$. 
Thus, the verifier can incrementally maintain the quantity
$\tilde{E}_i(u, r)$ in a streaming manner using space $O(\log |\field|)$:
while processing
the $i$th stream update,
the verifier simply
adds $\tilde{\delta}_{(u_i, v_i)}(u, r)+\tilde{\delta}_{(v_i, u_i)}(u, r)$ to the running sum
tracking $\tilde{E}_i(u, r)$.

\medskip
\noindent \textbf{Completeness.} It is evident that if the prover sends the true polynomial $g(Z)$,
then the verifier's check will pass, and the verifier will output the correct number of triangles in 
the final graph $G$.

\medskip
\noindent \textbf{Soundness.} If the prover sends a polynomial $s(Z) \neq g(Z)$,
then with probability at least $1-2\nodes/|\field| \geq 1-1/(3n^2)$ over the
verifier's random choice of $r \in \mathbb{F}$, it will hold that
$s(r) \neq g(r)$. Hence, with probability at least $1-1/(3n^2) \geq 2/3$,
the verifier's check will fail and the verifier will
reject. 
\end{proof}
%\vspace{-2mm}
\noindent Several remarks regarding Theorem \ref{thm:triangles} are in order.
\begin{itemize}
%\vspace{-2mm}
\item \textbf{Verifier Time.} The verifier in the protocol of Theorem \ref{thm:triangles}
can process each stream update in constant time as follows.
On stream update $e_i = (u_i, v_i)$, the verifier 
must add $\tilde{\delta}_{(u_i, v_i)}(u, r) + \tilde{\delta}_{(v_i, u_i)}(u, r)$ to each of the $E_i(u, r)$ values.
However, it is straightforward to check
that $\tilde{\delta}_{(u_i, v_i)}(u, r) = 0$ for all $u \neq u_i$,
so the verifier need only update two quantities at time $i$:
$E_i(u_i, r)$ and $E_i(v_i, r)$. We explain how 
both of these updates can be computed in constant time.
It can be seen that 
$\tilde{\delta}_{(u_i, v_i)}(u_i, r) = \frac{\prod_{1 \leq v' \leq \nodes: v' \neq v_i}(r-v')}{\prod_{1 \leq v' \leq \nodes: v' \neq v_i}(v_i-v')}.$
The right hand side of this equation can be computed in $O(1)$ time
if the verifier maintains a pre-computed lookup table consisting of $O(n)$ field elements.
Specifically, for each $v \in [\nodes]$, it suffices for the verifier to maintain 
the quantities $q_1(v) := \prod_{1 \leq v' \leq \nodes: v' \neq v_i} (r-v')$
and $q_2(v) = \left(\prod_{1 \leq v' \leq \nodes: v' \neq v_i}(v_i-v')\right)^{-1}$.
All $O(n)$ of these quantities can be computed in pre-processing in total time $O(\nodes \log \nodes)$, where the $\log \nodes$ term is due to the time required to compute a multiplicative
inverse in the field $\field$. Indeed, $q_1(1)$ and $q_2(1)$ can be computed naively in $O(n)$ time, and then for any $v > 1$,
$q_1(v)$ and $q_2(v)$ can be computed in $O(\log n)$ time from $q_1(v-1)$ and $q_2(v-1)$ via the identities
$q_1(v) = q_1(v-1) \cdot (r-v)^{-1} \cdot (r-v+1)$ and 
$q_2(v) = q_2(i-1) \cdot (v-1)^{-1} \cdot (n-v+1).$

Finally, the verifier can process the proof itself in time $O(n)$.
Indeed, recall that the proof consists of the values $s(x)$ for $x \in \{0, \dots, 2n\}$,
and the verifier simply needs to compute $\sum_{1 \leq x \leq n} s(x)$ as well
as $s(r)$. The first quantity can trivially be computed in time $O(n)$,
and the second can be computed in time $O(n)$ as well using standard
techniques (see, e.g., \cite{itcs}).
%\vspace{-2mm}
\item \textbf{Prover Time.}  The honest prover in the protocol of Theorem \ref{thm:triangles}
can be implemented to run in time $O(\length \cdot \nodes)$. 
Indeed,
the honest prover needs to evaluate $g(x)$ for $O(n)$ points $x \in \field$, and we have
explained above how $g(x)$ can be computed in $O(\length)$ time (in fact, in $O(1)$
time per stream update).
Note that this time is comparable to the cost of a naive triangle
counting algorithm that, for each edge and node combination, tests
whether the two edges incident on the edge and node exist in the graph. 

The time costs for both the prover and verifier are summarized in Table \ref{tab:trianglescost}.

\eat{
\item \textbf{More general streaming models.} The protocol of Theorem \ref{thm:triangles} can be trivially modified
to handle streams with deletions, as well as weighted and directed graphs.
In the case of weighted graphs, the weight assigned to a triangle $(u, v, z)$
is the \emph{product} of weights of the constituent edges $(u, v)$, $(v, z)$, and $(z, u)$.
}

%\vspace{-2mm}
\item \textbf{MA communication.} Theorem \ref{thm:triangles} implies that the (online) MA communication complexity of counting triangles is $O(n \log n)$
(see Section \ref{sec:commmodels} for the definition of the (online) MA communication model).
This essentially matches an $\Omega(n)$ lower bound on the (even non-online)  MA communication
complexity of the problem, proved by Chakrabarti \etal \cite{icalp} via 
a standard reduction to \textsc{set-disjointness}, and answers a question of
Cormode \cite{openproblem}.
%\vspace{-2mm}
\item \textbf{Extensions: Counting Structures Other Than Triangles.}
Let $H$ be a graph on $k$ vertices. It is possible to extend the protocol underlying Theorem \ref{thm:triangles} to count the number of occurrences of $H$ as a subgraph of $G$. The protocol requires $k - 2$ rounds, and its help and space costs are $O(k^3 n \log n)$ and $O(kn\log n)$ respectively. Appendix \ref{app:extend} has details.
\end{itemize}

\begin{table}
\centering
{\small
\begin{tabular}{|c|c|c|c|}
\hline
Verifier Pre-Processing & Verifier Time Per  & Verifier Time to & Prover Total\\ 
Time & 				Stream Update   & Process Proof & Time\\
\hline 
$O(n \log n)$ & $O(1)$ & $O(n)$ & $O(m \cdot n)$\\
\hline
\end{tabular}
\caption{Statement of Time Costs For Our \triangles Scheme (Theorem \ref{thm:triangles}).}
\label{tab:trianglescost}
}
\end{table}
%\vspace{-1mm}

%\vspace{-4mm}
\subsubsection{Comparison to Prior Work}
%\vspace{-1mm}
\label{sec:trianglescomparison}
As is typical in the literature on interactive proofs, the verifier 
in our \triangles\ protocol evaluates $g_{\stream}(r)$ for a random point
$r \in \field$, where $g$ is a polynomial derived from the input stream $\stream$. 
%This quantity serves as a ``secret'' that the verifier can use to 
%catch a dishonest prover.
However, our protocol qualitatively departs from prior work
in that $g_{\sigma}(r)$ is not a linear sketch of the input.
Here we define a linear sketch as any summary of the form $\mathbf{v} \in \field^{w}$ 
for some $w > 0$
which can be computed as $\mathbf{v} = S \mathbf{f}(\stream)$. Here, $S \in \field^{w\times n}$ is a ``sketch matrix'' and $\mathbf{f}(\stream)$ denotes the \emph{frequency-vector} of the stream, i.e., the $i$th element of $\mathbf{f}(\stream)$ is the number of times item $i$ appears in $\stream$. To the best of our knowledge, all existing protocols in the interactive proofs literature 
only require the verifier to compute a linear sketch of the input \cite{gkr,
shamir92, lfkn, icalp, vldb, itcs, gur, aw}. Typically, this linear sketch
consists of one or more evaluations of a \emph{low-degree extension} of the frequency vector $\mathbf{f}$ itself.

In contrast, 
in our \triangles\ protocol, we view the quantity $g_{i, \stream}(r) := \sum_{j \leq i} \Delta_j \cdot \tilde{E}_j(u_j, r) \cdot \tilde{E}_j(v_j, r)$ as the verifier's sketch at 
time $i$.
While $\tilde{E}_j(u, r)$ is a linear sketch of the input for each $j$ and node $u \in [\nodes]$
(in fact, this is what enables the verifier to compute $\tilde{E}_j(u, r)$ for each $u \in [\nodes]$ in a streaming
manner), $g_{i, \stream}(r) = \sum_{j \leq i} \Delta_j \cdot \tilde{E}_j(u_j, r) \cdot \tilde{E}_j(v_j, r)$ is not. A consequence is that,
in our \triangles\ protocol, %the verifier's update to her state at time $j$ depends
%on her state at time $j$ (specifically, her update depends on the stored values $\tilde{E}_j(u_j, r)$
%and $\tilde{E}_j(v_j, r)$). Moreover, 
the verifier's final state $g_{\length, \stream}(r)$ at the end of the data stream \emph{depends on the order of the stream stream tokens} --- it is easy to construct dynamic graph streams $\sigma, \sigma'$, such that $\sigma$ is a permutation of $\sigma'$, and yet $g_{\length, \stream}(r) \neq g_{\length, \stream'}(r)$ with high 
probability over the random choice of $r$. 

This contrasts with linear sketches, as %in
%a linear sketch, %each stream update $(i, \delta)$ contributes the value $S e_i$
%to the sketch independently of all other stream updates, where $e_i$ is the
%vector consisting of a 1 in the $i$th coordinate and zeros elsewhere. In particular,
the final state of any linear sketch is independent of the order of the data stream.
We conjecture that \emph{any}
semi-streaming scheme for counting triangles
must have a ``non-commuting'' verifier, as in the protocol of Theorem \ref{thm:triangles}.
In contrast, recent work has shown that, in the standard streaming model,
any ``non-commutative'' streaming algorithm that works
in the turnstile streaming model can be simulated with essentially
no increase space usage by a 
a linear sketching algorithm \cite{woodruffetal}.
%%\vspace{-mm}

\subsection{A Semi-Streaming Scheme for Maximum Matching}
We give a semi-streaming scheme for the \mm\ problem in general graphs. 
Our scheme combines the Tutte-Berge formula with algebraic techniques 
to allow the prover to prove matching upper and lower bounds on the size of a maximum matching
in the input graph.\footnote{It is possible to modify our scheme to give meaningful
answers on graphs with edges of negative
multiplicity. Specifically, the modified scheme can treat edges of negative multiplicity as having strictly positive multiplicity.
We omit the details for brevity.}

\begin{theorem} \label{thm:maxmatching-overview}[Formal Version of Theorem \ref{thm:mmoverview}] 
\label{thm:mm}
Assume there is an a priori upper bound $B \leq \poly(n)$ on the multiplicity of any edge in $G$. 
  There is an online scheme for \mm\ of total cost $O(B \cdot n \log n)$.
  Every scheme for \mm\ (online or prescient) requires
  the product of the space and help costs to be $\Omega(n^2)$, and hence requires total cost $\Omega(n)$, even for $B=1$.
\end{theorem}
\begin{proof}
The lower bound follows from an identical lower bound proved for the Bipartite Perfect Matching problem by Chakrabarti et al. \cite{icalp}, combined with the fact that the
\mm\ problem is at least as hard as Bipartite Perfect Matching. We now turn to the upper bound. 
We begin with a high-level proof sketch that highlights the novel aspects of our scheme,
before turning to a detailed scheme description.

\medskip
\noindent \textbf{Proof Sketch.}
The prover proves matching upper and lower bounds on the size of a maximum matching. To establish a lower bound, the prover simply sends a set of edges, $M$,
claimed to be a maximum matching, and uses techniques from prior work \cite{icalp} to establish that $M \subseteq E$. 
To establish an upper bound, we exploit the Tutte-Berge formula, which 
 states that the size of a maximum matching in $G=(V, E)$ is equal to
 \vspace{-1mm}
\begin{equation} \label{tutte} \frac12 \min_{U \subseteq V} \left(|U| - \odd(G-U) + |V|\right).\end{equation}
\vspace{-1mm}
Here, $G-U$ is the induced graph obtained from $G$ by removing all vertices in $U$, and $\odd(G-U)$ denotes the number of connected
components with an odd number of vertices in $G-U$. The prover sends a set $U^* \subseteq V$ claimed to achieve the minimum in 
Equation \eqref{tutte}. The primary novelty in our scheme is a method for computing $\odd(G-U^*)$ that (in contrast to the \mm\ scheme of \cite{esa})
avoids the need for the prover to replay all of the edges that appeared in the stream. 

\medskip \noindent \textbf{Detailed Scheme Description.}
Suppose that the prover claims that the answer is $k$. Our scheme consists of two parts: in the first part, the prover proves that
the size of a maximum matching in $G$ is \emph{at least} $k$. In the second part, the prover proves that the size of a maximum matching in $G$ is \emph{at most} $k$.
We clarify that the verifier will be able to perform the required processing for both parts of the scheme simultaneously with a single pass over the input stream,
and the prover can send the annotation for both parts of the scheme in a single message (i.e., by simply concatenating the annotation for the two parts). 
If the prover successfully passes all of the verifier's checks in both parts, then the verifier
is convinced that the size of a maximum matching in $G$ is exactly $k$.

\medskip 
\noindent \textbf{Part One.} 
This part of the scheme is similar to the analogous part of the bipartite perfect matching protocol given in \cite{icalp}: 
the prover sends a set $M$ of $k$
edges $M=(e^{(1)}, \dots, e^{(k)})$ that are claimed to comprise a valid matching in $G$, and the verifier
explicitly stores $M$ (this requires at most $n \log n$ bits of annotation and space). 
The verifier needs to check that $M$ is indeed a valid matching for $G$.
This requires checking two properties:

\begin{itemize}
\item \textbf{Property 1:} For all nodes $v \in V$, $v$ is incident to at most one edge $e^{(i)} \in M$. 
 \item \textbf{Property 2:} $M \subseteq E$. 
\end{itemize}

Property 1 is trivial to check, because the verifier stores $M$ explicitly. 
Property 2 can be checked using a scheme for the \mysubset\ problem described in \cite[Lemma 5.3]{icalp}. Here, the input
to the \mysubset\ problem consists of two (multi-)sets, $S_1$ and $S_2$, over a data universe $\mathcal{U}$, arbitrarily interleaved, and 
the output is 1 if and only if $S_1 \subseteq S_2$. We apply the \mysubset\ scheme of Chakrabarti et al. with $S_1 = M$, $S_2 = E$, and $\mathcal{U}$ equal
to the set of all $O(n^2)$ possible edges.
The scheme requires help cost $O(Bx \log n)$ and space cost $O(By \log n)$ for any $x \cdot y \geq |\mathcal{U}| = O(n^2)$, where
$B$ is an a prior upper bound on the multiplicity of any edge in $E$. In particular, assuming that $B=O(1)$, both the help and space
costs of the \mysubset\ scheme can be set to $O(n \log n)$.

\medskip 
\noindent \textbf{Part Two.} 
We exploit the Tutte-Berge formula. This formula states that the size of a maximum matching in a graph $G=(V, E)$ is equal to

\[\frac12 \min_{U \subseteq V} \left(|U| - \odd(G-U) + |V|\right).\]

Here, $G-U$ is the induced graph obtained from $G$ by removing all vertices in $U$, and $\odd(G-U)$ denotes the number of connected
components with an odd number of vertices in $G-U$. 

Thus, to establish that the size of the a maximum matching in $G$ is at most $k$, the prover first identifies a subset $U^* \subseteq V$
such that $k=g(U^*)$, where $$g(U^*) := \frac12 \left(|U^*| - \odd(G-U^*) + |V|\right).$$ 
The prover sends $U^*$ to $V$, which requires $O(n \log n)$ bits of help, and the verifier stores $U^*$ explicitly, which requires $O(n \log n)$ 
bits of space.
Clearly the verifier can easily compute the quantity $g(U^*)$ if she knows the value $\odd(G-U^*)$. The remainder of the scheme
is therefore devoted to computing $\odd(G-U^*)$. 

\medskip 
\noindent \textbf{A (Sub-)Scheme for Computing $\odd(G-U^*)$.} The prover first assigns each connected component in $G-U^*$ 
a unique label in $[C]$ arbitrarily, where $C$ is the number of connected components in $G-U^*$. 
The prover then sends a list $L$ to the verifier that contains a pair $(v, \ell_v)$ for each node $v \in V \setminus U^*$, 
where $\ell_v$ is claimed to equal the label of $v$'s connected
component in $G-U^*$. The verifier stores the list $L$ explicitly --- the help and space costs for the prover to send  $L$ and 
the verifier to store $L$ are both $O(n \log n)$. Since the verifier stores $L$ and $U^*$ explicitly, it is trivial
for the verifier to check that  every node in $V \setminus U^*$ appears exactly once in $L$.

If the list $L$ is as claimed, then it is easy for the verifier to compute $\odd(G-U^*)$ in $O(n \log n)$ space,
as the verifier can simply count of the number of labels that 
appear in $L$ an odd number of times. The remainder of the scheme
is therefore devoted to ensuring that the list $L$ is as claimed. 

To this end, the verifier must ensure that the labels $\ell_v$ in $L$ actually correspond to the connected components of $G-U^*$.
This requires checking two properties. 

\begin{itemize}
\item \textbf{Property A:} For each label $\ell$, all nodes $v$ with label $\ell_v=\ell$ are connected to each other in $G-U^*$. 
\item \textbf{Property B:} For every pair of nodes $u, v$ in $V \setminus U^*$ with different labels $\ell_u \neq \ell_v$,
it holds that $(u, v) \not\in E$. That is,
there is no edge in $G-U^*$ connecting two nodes that are claimed to be in different connected components.
\end{itemize}

To prove that Property A holds, the prover sends, for each connected component $\ell$ in $G-U^*$, a set of edges $T_{\ell} \subseteq \left(V\setminus U^*\right) \times \left(V\setminus U^*\right) $ that is claimed to be a spanning tree for all
nodes with label $\ell$
(and the verifier explicitly stores each of the $T_{\ell}$'s).
Note that sending and storing the $T_\ell$'s requires $O(n \log n)$ bits of help and space in total. 
For each $\ell$, the verifier must check that $T_\ell$ spans  all nodes with label $\ell$ in $L$,
and that $\cup_{\ell} T_\ell \subseteq E$. 

Checking that $T_{\ell}$ spans all nodes with label $\ell$ is trivial, as the verifier has explicitly stored the list $L$ of (vertex, label) pairs, as well as the trees $T_{\ell}$. 
To check that $\cup_{\ell} T_{\ell} \subseteq E$, we use the \mysubset\ scheme of \cite[Lemma 5.3]{icalp}.

\medskip 
\noindent \textbf{A (Sub-)Scheme for Checking that Property B holds.} 
Checking that Property B holds requires more care.
Notice that Property B holds if and only if:

\begin{equation} \label{eq:thebigeq} 0=\sum_{(u, v) \in [n] \times [n] } D(u, v) \cdot E(u, v). \end{equation}

Here, $D(u, v) \colon [n] \times [n] \rightarrow \{0, 1\}$ is defined to be the function that outputs 1 if $\ell_u \neq \ell_v$, and is 0 otherwise (to clarify, if either $u$ or $v$ is in $U^*$ itself, then $D(u, v)$ is defined to be 0) --- we choose the letter $D$ because such edges are \emph{disallowed}, if the component labels provided by the prover are correct. 
Abusing notation, $E(u, v) \colon [n] \times [n] \rightarrow \{0, 1\}$ in Equation \eqref{eq:thebigeq} denotes the function that on input $(u, v)$, outputs the multiplicity of edge $(u, v)$ in $E$. 

To check that Equation \eqref{eq:thebigeq} holds, we let $\tilde{D}$ denote the unique bivariate polynomial over a field $\mathbb{F}$ of prime order, $2n^3 \leq |\mathbb{F}| \leq 4 n^3$, such
that the degree of $\tilde{D}$ is at most $n$ in each variable, and $\tilde{D}(u, v) = D(u, v)$ for all $(u, v) \in [n] \times [n]$. Likewise, we let 
$\tilde{E}$ denote the unique bivariate polynomial over $\mathbb{F}$ such that the degree of $\tilde{E}$ is at most $n$ in each variable, and 
$\tilde{E}(u, v) = E(u, v)$ for all $(u, v) \in [n] \times [n]$. 

\medskip 
\noindent \textbf{Prover's Computation.}
After the stream has passed and the prover has sent the list $L$, 
the prover is required to send a univariate polynomial $s(Y)$ of degree at most $2n$ claimed to equal

\[g(Y) := \sum_{u \in [n]} \tilde{D}(u, Y) \cdot \tilde{E}(u, Y).\]

Because of the degree bound on $s(Y)$, this can be done with help cost $O(n \log |\mathbb{F}|) = O(n \log n)$. 

\medskip 
\noindent \textbf{Verifier's Computation.}
At the start of the stream, the verifier picks a random $r \in \mathbb{F}$. 
We will explain in the next two paragraphs how the verifier can evaluate $g(r)$ in a streaming fashion, with space cost $O(n \log n)$.
The verifier then checks that $s(r)= g(r)$. If this check fails, then the verifier halts and rejects. If the check passes, then verifier
checks that $\sum_{v=1}^n s(i)=0$. If so, the verifier is convinced that Property B holds, and
accepts $k$ as an upper bound on the size of any maximum matching in $G$. 

While observing the data stream, the verifier incrementally computes
the $n$ values $\tilde{E}(u, r)$ for each $u \in [n]$. The proof of Theorem \ref{thm:triangles} explained 
how the verifier can do this in a streaming manner, using $O(\log |\mathbb{F}|) = O(\log n)$ space for each value $u \in [n]$,
and thus $O(n \log n)$ space in total. 
Similarly, after the prover has sent $U^*$ and $L$, the verifier computes the $n$ values $\tilde{D}(u, r)$ for each $u \in [n]$. 
Notice that the function $\tilde{D}$ is uniquely determined by $U^*$ and the list $L$ of (vertex, label) pairs. Since
the verifier has explicitly stored $U^*$ and $L$, it is straightforward for the verifier to evaluate each of these $n$ values with
total space $O(n \log|\mathbb{F}|) = O(n \log n)$ (in the remarks following
the theorem, we also explain in detail how the verifier can perform these
$n$ evaluations in $O(n)$ total time).

Once the verifier has computed the values $\tilde{E}(u, r)$  and $\tilde{D}(u, r)$ for each $u \in [n]$, it 
is straightforward for the verifier to compute $g(r) = \sum_{u \in [n]} \tilde{E}(u, r) \cdot \tilde{D}(u, r)$. 

\medskip \noindent \textbf{Soundness and Completeness of the (Sub-)Scheme for Property B.} The proofs of soundness and completeness are 
essentially identical to Theorem \ref{thm:triangles}. If Equation \eqref{eq:thebigeq} holds and $s=g$, then the verifier's checks will pass with probability 1, and 
the verifier will accept $k$ as an upper bound on the size of any maximum matching in $G$. If Equation \eqref{eq:thebigeq} is false,
then the prover is forced to send a polynomial $s \neq g$, or the verifier's final check that $\sum_{v=1}^n s(v) =0$ will fail. But if $s \neq g$, then
$s(r) \neq g(r)$ with probability at least $1-2n/|\mathbb{F}| \geq 1-1/n$, because any two distinct polynomials
of degree at most $2n$ can agree on at most $2n$ inputs.
\end{proof}

\

Several remarks are in order regarding Theorem \ref{thm:mm}.

\begin{itemize}
\item \textbf{The Bottleneck in Reducing Space Costs.} In a formal sense, the most difficult part of the \mm\ scheme given in Theorem \ref{thm:mm} is checking that Property B holds. 
Indeed, this is the only part of the entire \mm\ scheme
for which we are presently unable to reduce the verifier's space usage to $o(n)$ without increasing the help cost to $\Omega(n^2)$.
That is, if it were possible for the verifier to \emph{avoid} checking Property B, we would be able to give a scheme with help cost $O(x \log n)$ and space cost $O(y \log n)$
for any pair $x, y$ such that $x \cdot y \geq n^2$ and $x \geq n$ (in contrast to Theorem \ref{thm:mm}, which only gives a scheme for the special values $x=y=n$). 
We conjecture, however, that the difficulty is inherent, i.e., that there is no way to reduce the space cost to $o(n)$ without increasing the help cost to $\Omega(n^2)$.
 
 \item \textbf{Prover Time.} In the scheme of Theorem \ref{thm:mm}, the prover has to compute a maximum matching and 
 identify the (possibly non-unique) set $U^*$ whose existence is guaranteed by the Tutte-Berge formula. Fortunately, standard algorithms for computing maximum matchings identify the set $U^*$ as a natural byproduct of their execution.\footnote{See for example the Boost Graph Library, \url{http://www.boost.org/doc/libs/1_45_0/libs/graph/doc/maximum_matching.html.}}
 Using Fast Fourier Transform techniques described in \cite{itcs}, the remainder of the prover's computation in the scheme 
 of Theorem \ref{thm:mm} can be executed in time $O(n^2 \log n)$. Hence, the total runtime of the prover is $O(T + n^2 \log n)$,
 where $T$ is the time required to find a maximum matching along with the Tutte-Berge decomposition $U^*$.
 
 \item \textbf{Verifier Time.} Recall that we explained in Section \ref{sec:triangles} how to implement a verifier 
 running in constant time per stream update, after a pre-processing stage requiring $O(n)$ time (see the remarks following Theorem \ref{thm:triangles}). These techniques are easily modified to enable the verifier in the scheme of Theorem \ref{thm:mm}
 to also run in constant time per stream update, after a pre-processing stage requiring $O(n \log n)$ time. 
 
 Finally, we explain that the verifier can process the proof in $O(n)$ time. Indeed, when processing the proof, the verifier's tasks fall into three categories: (1) run two \mysubset\ protocols involving the sets $M$ and $\cup_{\ell} T_{\ell}$
 provided by the prover (note both sets are of size at most $n$), (2) Evaluate $\tilde{D}(u, r)$ for each $u \in [n]$ as part of the (sub-)scheme to check that Property B holds, and (3)
 perform various other checks on the proof that require $O(n)$ time in total by inspection. 
 
The verifier's computation for the two \mysubset\ schemes on the sets $M$ and $\cup_{\ell} T_{\ell}$, both of size at most $n$, can easily be implemented in $O(n)$
total time using the same techniques as above the ensure that the verifier processes each element of $M$ and $\cup_{\ell} T_{\ell}$ in constant time --- we omit the straightforward details
for brevity. Evaluating 
 $\tilde{D}(u, r)$ for each $u \in [n]$ in total time $O(n)$ can be done in a similar fashion, but requires more care. It is straightforward to check that 
 
 \begin{equation}
 \label{eq:finalmaxeq} \tilde{D}(u, r) = \sum_{v \in V \setminus U^*: \ell_{v} \neq \ell_u} \tilde{\delta}_{v}(r),\end{equation}
 where \[\tilde{\delta}_{v}(X) = \frac{\prod_{1 \leq v' \leq n, v \neq v'} (r-v')}{\prod_{1 \leq v' \leq n, v \neq v'}(v-v')}.\]
 
 Using the techniques described in Section \ref{sec:triangles}, the verifier can store and compute the values  $\tilde{\delta}_v(r)$ for all $v \in [n]$ in pre-processing,
 in $O(n \log n)$ total time. The verifier also computes and stores the value $H:=\sum_{v \in V \setminus U^*}\tilde{\delta}_{v}(r)$ in $O(n)$ time.
 
 With these values in hand, the verifier can compute all $n$ values $\tilde{D}(u, r): u \in [n]$ in $O(n)$ total time. To see this, notice that
 Equation \eqref{eq:finalmaxeq} implies that $\tilde{D}(u, r)$ depends only on the label $\ell_u$ of the connected component of $u$ in $G-U^*$. That is, $\tilde{D}(u, r) = \tilde{D}(u', r)$
 for all $u'$ such that $\ell_u = \ell_u'$. Moreover, letting $S(\ell)$ denote the set of all vertices $v$ with $\ell_v=\ell$, $\tilde{D}(u, r)$ can be computed in time 
 $O(|S(\ell_u)|)$, given the precomputed values described above, via the following identity that follows from Equation \eqref{eq:finalmaxeq}: 
 $$\tilde{D}(u, r) = H - \sum_{v: \ell_v = \ell_u}\tilde{\delta}_{v}(r).$$
 
 Hence, \emph{all} of the $\tilde{D}(u, r)$ values can be computed in total time $\sum_{\text{distinct labels } \ell_u} O(|S(\ell_u)|) = O(n)$.
 
 The prover and verifier's time costs in the \mm\ scheme of Theorem \ref{thm:mm} are summarized in Table \ref{tab:mmcost}.
 
 \end{itemize}
 
 \begin{table}
\centering
\begin{tabular}{|c|c|c|c|}
\hline
Verfier Pre-Processing & Verifier Time Per  & Verifier Time to & Prover Total\\ 
Time & 				Stream Update   & Process Proof & Time\\
\hline 
$O(n \log n)$ & $O(1)$ & $O(n)$ & $O(T + n^2 \log n)$\\
\hline
\end{tabular}
\caption{Statement of time costs for our \mm\ scheme (Theorem \ref{thm:mm}). $T$ denotes
the time required to find a maximum matching, as well as the Tutte-Berge decomposition of $G$.}
\label{tab:mmcost}
\end{table}

\section{Lower Bounds for \connectivity\ and \bipartiteness}
\label{sec:lb}
In this section, we establish our lower bounds on the cost of online schemes
for \connectivity\ and \bipartiteness\ in the XOR update models. 
Like almost all previous lower bounds for data stream computations, our lower bounds use reductions from problems in communication complexity. 
To model the prover in a scheme, the appropriate communication setting is Merlin-Arthur communication, which we now introduce.

\subsection{Merlin-Arthur Communication}
\label{sec:commmodels}
Consider a communication game involving three parties, named Alice, Bob, and Merlin. 
Alice holds an input $x \in \mathcal{X}$, Bob and input $y \in \mathcal{Y}$,
and Merlin is omniscient (he sees both $x$ and $y$) but untrusted. 
Alice and Bob's goal is to compute $f(x, y)$ for some agreed upon function $f: \mathcal{X} \times \mathcal{Y} \rightarrow \{0, 1\}$.

In an MA communication protocol $\mathcal{P}$, Merlin first broadcasts a message $m_M$ to both Alice and Bob. Alice and Bob then engage in 
a randomized communication protocol, before outputting a single bit. To clarify, Merlin does not learn the randomness that Alice and Bob
use until after sending the message $m_M(x, y)$. For each input $(x, y)$, 
the protocol $\mathcal{P}$ defines a game between Merlin, Alice, and Bob, in which Merlin's goal is to make Alice and Bob output 1. 
We define the value $\textbf{Val}^{\mathcal{P}}(x, y)$ to be Merlin's probability of winning this game with optimal play. 
Given a Boolean function $f$, we say that $\mathcal{P}$ computes $f$ if, for all $(x, y)$ we have
(1) $f(x,y)=0 \Longrightarrow  \textbf{Val}^{\mathcal{P}}(x, y) \leq  1/3$, and (2) $f(x,y)=1 \Longrightarrow \textbf{Val}^{\mathcal{P}}(x, y) \geq 2/3$. 
We refer to the Property (1) as \emph{soundness} and Property (2) as \emph{completeness}.
Notice that the completeness requirement is qualitatively weaker than what we require for \emph{schemes}:
in a scheme, we required that there exist a prover strategy that can convince the verifier of the value of $f(x)$ for \emph{all} $x$,
while in an MA communication protocol we only require this to hold for $x \in f^{-1}(1)$.

The \emph{help cost}, or $\hcost(\cP)$, of $\mathcal{P}$ is $\max_{(x, y)} |m_M(x, y)|$, i.e., the maximum length of Merlin's message in bits. 
The \emph{verification cost}, or $\vcost(\cP)$, of $\mathcal{P}$ is the maximum number of bits that Alice and Bob exchange, where the maximum is taken over
all inputs $(x, y)$, all possible Merlin messages $m_M$, and all choices of Alice and Bob's randomness. The \emph{total cost} of $\mathcal{P}$
is the sum of the help and verification costs of $\mathcal{P}$. 

In an \emph{online} MA communication protocol (OMA protocol for short), neither Merlin nor Bob can talk to Alice. 
This condition models the one-pass streaming restriction on the verifier in an online scheme. 
Indeed, given any online scheme for a function $f$, we naturally obtain an OMA protocol $\cP$
for the communication problem in which Alice holds a prefix of a stream, Bob holds a suffix, and 
the goal is to evaluate $f$ on the concatenated stream $x \circ y$. The help cost of $\cP$ is equal to the help cost of the scheme,
while the verification cost of $\cP$ is equal to the space cost of the scheme. Hence, if we establish
lower bounds on the help and verification costs of any OMA protocol for $f$, we may conclude an
equivalent lower bound on the help and space costs of any online scheme for $f$. 

In this section, we establish lower bounds on the help and verification costs of any OMA protocol
for the \disconnectivity\ and \bipartiteness\ problems in the XOR update model.
 More precisely,
we consider the communication problems \disconnectivitycc\ and \bipartitenesscc\
in which Alice holds the first $\length - n$ tuples in a graph stream in the XOR update model,
Bob holds the length $n$ tuples, and the output function evaluates to 1 if and only if the resulting graph 
is disconnected or bipartite, respectively.\footnote{The reason that
we consider \disconnectivitycc\ rather than \connectivitycc\ is 
the asymmetric way that inputs
in $F^{-1}(0)$ and $F^{-1}(1)$ in the definition of OMA communication complexity.
Recall that the OMA communication problem for a decision problem $F$ requires only that if $F(x) = 1$ then there is some prover that will cause the verifier to accept with high probability, and if $F(x) = 0$ then there is no such prover. (By contrast, our definition of a scheme for a function $F$ requires there to be a convincing proof of the value of $F(x)$ for all values $F(x)$.) Hence, the OMA communication complexities of \disconnectivitycc\ and \connectivitycc\ may not be equal, and indeed our lower bound argument applies only to \disconnectivitycc.}

\subsection{The Lower Bound}

\begin{theorem}
\label{thm:lb}
Consider any OMA protocol $\mathcal{P}$ for \disconnectivitycc or \bipartitenesscc. Then 
$$(\hcost(\cP)+n) \cdot \vcost(\cP) = \Omega(n^2).$$
This holds even under the promise that the first $\length - n$ stream updates (i.e., Alice's input) are all unique, and the last $n$ stream updates (i.e., Bob's input) are all incident to a single node. 
In particular, the total cost of $\mathcal{P}$ is $\Omega(n)$. 
\end{theorem}
\begin{proof}
We begin with the \disconnectivitycc problem. Let $\mathcal{P}$ denote any OMA protocol for \disconnectivitycc that 
works on graphs with $n+1$ nodes,
under the promise described in the theorem hypothesis.
As discussed in the outline of Section \ref{sec:lboverview}, our proof will use a reduction from the \idx\ problem on ${n \choose 2}$ inputs.
In this problem, Alice's input consists of a bitstring $\bx$ of length ${n \choose 2}$, Bob's input is an index $i^* \in [{n \choose 2}]$, 
and the goal is to output $\bx_{i^*}$. It was established in \cite{icalp} that any OMA protocol $\mathcal{Q}$ for \idx\ on ${n \choose 2}$ inputs
requires \begin{equation}
\label{eq:lbeq} \hcost(\cQ) \cdot \vcost(\cQ) = \Omega(n^2).\end{equation}
We show how to use $\cP$ to construct a protocol $\cQ$ for \idx\ on  ${n \choose 2}$ inputs with $\hcost(\cQ) = n + \hcost(\cP)$ and $\vcost(\cQ) = \vcost(\cP)$. 
It will then follow from Equation \eqref{eq:lbeq} that  $(\hcost(\cP)+n) \cdot \vcost(\cP) \geq n^2$ as claimed.

\medskip
\noindent \textbf{Description of $\cQ$.} 
In $\cQ$, Alice interprets her input $\bx \in \{0,1\}^{{n \choose 2}}$ as an undirected graph $G_1$ with $n$ nodes as follows. She associates 
each index $i \in {n \choose 2}$ with a unique edge $(u_i, v_i)$ out of the set of all ${n \choose 2}$ possible edges that could appear in $G_1$.
 Alice also adds to $G_1$ a special node $v^*$, and Alice connects $v^*$ to every other node $v$ in $G_1$. Denote
 the resulting graph on $n+1$ nodes by $G_2$. Notice that $G_2$ is always connected
 (as every node is connected to $v^*$ by design). %Alice now runs her part of the \connectivity$^\text{cc}$ protocol $\cP$, sending a message to Bob. 

Likewise, Bob interprets his input $i^* \in [{n \choose 2}]$ as an edge $(u_{i^*}, v_{i^*})$. Clearly,
determining whether $\bx_{i^*}=1$ is equivalent to determining whether edge $(u_{i^*}, v_{i^*})$ appears in Alice's graph $G_2$. 
Merlin sends Bob a list $L$ claimed to equal all edges incident to node $u_{i^*}$ in $G_2$. This requires only $n$ bits of ``help'', since there are only $n$ nodes to which $u_{i^*}$ might be adjacent. 
Bob treats $L$ as his input to the \disconnectivitycc problem. 

Alice, Bob, and Merlin now run the \disconnectivitycc\ protocol $\cP$ (with Alice's input equal to $G_2$ and Bob's input equal to $L$).
Bob outputs 1 if and only if the protocol $\cP$ outputs 1, and $L$ contains the edge $(u_{i^*}, v_{i^*})$.

\medskip
\noindent \textbf{Costs of $\cQ$.} The help cost of $\cQ$ is equal to $n + \hcost(\cP)$, since the honest Merlin sends Bob the list $L$, and 
then behaves as he would in the protocol $\cP$. The verification cost of $\cQ$ is just $\vcost(\cQ)$, since the only message Alice sends to Bob 
is the message she would send in $\cP$.

\medskip
\noindent \textbf{Completeness and Soundness of $\cQ$.}
Let $G_3$ denote the graph obtained from $G_2$ by XORing all the edges in the list $L$. Let $I(u_{i^*})$ 
denote the set of edges incident to $u_{i^*}$ in $G_3$. 
We claim that $G_3$ is disconnected if and only if $L$ is equal to $I(u_{i^*})$. 
For the first direction, suppose that $L$ is equal to $I(u_{i^*})$. Then by XORing the edges
in $G_3$ with the edges in $L$, every edge incident to node $u_{i^*}$ is deleted from the graph. Hence, $u_{i^*}$ is an isolated
vertex in $G_3$, implying that $G_3$ is disconnected. 

For the second direction, suppose that $L$ is not equal to $I(u_{i^*})$. Let $(u_{i^*}, v)$
denote an edge in $L \setminus I(u_{i^*})$. Then $(u_{i^*}, v)$ is in the graph $G_3$. Moreover, $v$ is adjacent
to node $v^*$, as are all nodes in $G_3$ other than $u_{i^*}$. Hence $G_3$ is connected.

To complete the proof of completeness of $\cQ$, note that if $\mathbf{x}_{i^*}=1$, then the edge $(u_{i^*}, v_{i^*})$ is in $G_3$.
If Merlin sends $L=I(u_{i^*})$, then $G_3$ will be disconnected, and by the 
 the completeness of $\cP$, Merlin can convince Bob that $G_3$ is disconnected with probability at least $2/3$. In this event,
 Bob will output 1, because $(u_{i^*}, v_{i^*})$ will be in the list $L$.

To complete the proof of soundness of $\cQ$, note that if $\mathbf{x}_{i^*}=0$, then the edge $(u_{i^*}, v_{i^*})$ is not in $G_3$.
Hence, if Merlin sends $L=I(u_{i^*})$, then Bob will reject automatically, because  $(u_{i^*}, v_{i^*})$  will not be in the list $L$. 
On the other hand, if Merlin sends a list $L$ that is \emph{not} equal to $I(u_{i^*})$, then $G_3$ will be connected. By the 
 the soundness of $\cP$, Merlin can convince Bob that $G_3$ is disconnected with probability at most $1/3$. Hence,
 Bob will output 1 in $\cQ$ with probability at most $1/3$, completing the proof for \disconnectivitycc.

\medskip
 
 \noindent \textbf{Proof for \bipartitenesscc.} The proof for the \bipartitenesscc\ problem follows a similar high-level outline. 
 Let $\cP'$ be an online MA protocol for \bipartitenesscc\ on graphs with $n+1$ nodes. 
 We show how to use $\cP'$ to 
 to construct a protocol $\cQ'$ for \idx\ on  $(n/2) \cdot (n/2)$ inputs with $\hcost(\cQ') = n + \hcost(\cP')$ and $\vcost(\cQ') = \vcost(\cP')$ (we assume that $n$ is even for simplicity). 
It will then follow from Equation \eqref{eq:lbeq} that  $(\hcost(\cP')+n) \cdot \vcost(\cP') = \Omega(n^2)$ as claimed.

\medskip
\noindent \textbf{Description of $\cQ'$.} 
In $\cQ'$, Alice interprets her input $\bx \in \{0,1\}^{{n^2 / 4}}$ as an undirected bipartite graph $G'_1$ with $n/2$ nodes on the left and $n/2$ nodes on the right, as follows. She associates 
each index $i \in [n^2/4]$ with a unique edge $(u_i, v_i)$ out of the set of all $n^2/4$ possible edges that could appear in the bipartite graph $G'_1$.
 Alice also adds to $G'_1$ a special node $v^*$, and Alice connects $v^*$ to all $n/2$ nodes on the right side of $G'_1$. Denote
 the resulting graph on $n+1$ nodes by $G'_2$. Notice that $G'_2$ is always bipartite by design. Indeed, letting $S_1$ denote the set of all nodes on the left side of $G'_1$, and $S_2$ denote the set of all nodes on the right side of $G_1$, then every edge in $G'_2$ connects a vertex in $S_1 \cup \{v^*\}$ to one in $S_2$.
 
 %Alice now runs her part of the \connectivity$^\text{cc}$ protocol $\cP'$, sending a message to Bob. 

Likewise, Bob interprets his input $i^* \in [n^2/4]$ as an edge $(u_{i^*}, v_{i^*})$. Clearly,
determining whether $\bx_{i^*}=1$ is equivalent to determining whether edge $(u_{i^*}, v_{i^*})$ appears in Alice's graph $G'_2$. 
Merlin sends Bob a list $L$ claimed to equal all edges incident to node $u_{i^*}$ in $G'_2$. This only requires $n$ bits of ``help'', since there are only $n$ nodes that $u_{i^*}$ might be adjacent to. 
Bob treats $L \cup \{(u_{i^*}, v^*)\}$ as his input to the \bipartitenesscc problem. 

Alice, Bob, and Merlin now run the \bipartitenesscc protocol $\cP'$ (with Alice's input equal to $G'_2$ and Bob's input equal to $L \cup \{(u_{i^*}, v^*)\}$).
Bob outputs 1 if and only if the protocol $\cP'$ outputs 1, and $L$ contains the edge $(u_{i^*}, v_{i^*})$.

\medskip
\noindent \textbf{Costs of $\cQ'$.} Analogous to the case of \disconnectivitycc, the help cost of $\cQ'$ is equal to $n + \hcost(\cP')$, since the honest Merlin sends Bob the list $L$, and 
then behaves as he would in the protocol $\cP'$. The verification cost of $\cQ'$ is just $\vcost(\cQ')$, since the only message Alice sends to Bob 
is the message she would send in $\cP'$.

\medskip
\noindent \textbf{Completeness and Soundness of $\cQ'$.}
Let $G'_3$ denote the graph obtained from $G'_2$ by XORing all the edges in $L \cup \{(u_{i^*}, v^*)\}$. Let $I(u_{i^*})$ 
denote the set of edges incident to $u_{i^*}$ in $G'_3$. 
We claim that $G'_3$ is bipartite if and only if $L$ is equal to $I(u_{i^*})$. 
For the first direction, suppose that $L$ is equal to $I(u_{i^*})$. Then by XORing the edges
in $G'_3$ with the edges in $L \cup \{(u_{i^*}, v^*)\}$, every edge incident to node $u_{i^*}$ in $G'_2$ is deleted from the graph, and edge $\{(u_{i^*}, v^*)\}$ is inserted. It follows that $G'_3$ is bipartite. Indeed, recall that $S_1$ denotes the set of all nodes on the left side of $G'_1$, and $S_2$ denotes the set of all nodes on the right side of $G'_1$. 
Then every edge in $G'_3$ connects a vertex in $\left(S_1 \setminus \{u_{i^*}\} \right)\cup \{v^*\}$ to a node in $S_2 \cup \{u_{i^*}\}$. That is, if $L$ is equal to $I(u_{i^*})$,
then the bipartition of $G'_3$ is identical to that of $G'_2$, except that $u_{i^*}$ has switched from the left hand side of $G'_2$ to the right hand side of $G'_3$.

For the second direction, suppose that $L$ is not equal to $I(u_{i^*})$. Let $(u_{i^*}, v)$
denote an edge in $L \setminus I(u_{i^*})$. Then $(u_{i^*}, v)$ is in the graph $G'_3$. Moreover, $v$ is adjacent
to node $v^*$ in $G'_3$, as is $u_{i^*}$. Hence, there is a triangle in $G'_3$, so $G'_3$ cannot be bipartite.

To complete the proof of completeness of $\cQ'$, note that if $\mathbf{x}_{i^*}=1$, then $(u_{i^*}, v_{i^*})$ is in $G'_3$.
If Merlin sends $L=I(u_{i^*})$, then $G'_3$ will be bipartite, and by the 
 the completeness of $\cP'$, Merlin can convince Bob that $G'_3$ is indeed
 bipartite with probability at least $2/3$. In this event,
 Bob will output 1 in $\cQ'$, because $(u_{i^*}, v_{i^*})$ will be in the list $L$.

To complete the proof of soundness of $\cQ'$, note that if $\mathbf{x}_{i^*}=0$, then $(u_{i^*}, v_{i^*})$ is not in $G'_3$.
Hence, if Merlin sends $L=I(u_{i^*})$, then Bob will reject automatically, because  $(u_{i^*}, v_{i^*})$  will not be in the list $L$. 
On the other hand, if Merlin sends a list $L$ that is \emph{not} equal to $I(u_{i^*})$, then $G'_3$ will be non-bipartite. By the 
 the soundness of $\cP'$, Merlin can convince Bob that $G'_3$ is bipartite with probability at most $1/3$. Hence,
 Bob will output 1 in $\cQ'$ with probability at most $1/3$, completing the proof for \bipartitenesscc.
\end{proof}

Because any online scheme for \connectivity\ or \bipartiteness\ can be simulated
by an OMA communication protocol, we obtain the following corollary.

\begin{corollary}[Formal Version of Theorem \ref{thm:connectivity}] 
\label{cor:lb}
Consider any online scheme for \connectivity\ or \bipartiteness\ in the XOR update model with help cost $\anncost$
and and space cost $\vercost$. Then $(\anncost+n) \cdot \vercost  \geq n^2$, 
even under the promise that the first $\length - n$ stream updates are all unique, and the last $n$ stream updates are all incident to a single node. In particular, the total cost of any annotation scheme for these problems is $\Omega(n)$. 
\end{corollary}

\section{Open Questions}
\eat{We put forth the notion of \emph{semi-streaming annotation schemes} for 
verifiably outsourcing problems on graph streams,
and gave evidence that semi-streaming annotation schemes represent a substantially
more robust solution concept than does the standard (sans prover) semi-streaming model.
Specifically, we gave new semi-streaming annotation schemes for two
basic graph problems (\triangles\ and \mm),
and we identified two graph problems (\connectivity\ and \bipartiteness\ in the XOR
edge update model) that do possess standard semi-streaming algorithms,
but do not possess schemes of ``sub-semi-streaming'' cost.}

A number of questions regarding annotated data streams in general,
and semi-streaming annotation schemes in particular, remain open. Here,
we highlight five.

\begin{itemize}

\item Exhibit any explicit function for which all online schemes require
total cost asymptotically larger than the square root of the input size.
In particular, it is open to exhibit any graph problem that \emph{cannot}
be solved by an online semi-streaming scheme.

\item Do there exist schemes of total cost $o(n^2)$ for any of the following graph problems:
shortest $s$-$t$ path in general graphs, diameter, and computing the value of a maximum flow. Note that
a semi-streaming scheme is known for shortest $s$-$t$ path in graphs of polylogarithmic diameter \cite{esa}, but not in general graphs (neither directed nor undirected).

\item Do there exist schemes of total cost $o(n)$ for connectivity or bipartiteness in
the strict turnstile update model? What about the insert-only update model? We conjecture that the answer is no in all cases. 

\item Is it possible to give a scheme for \triangles\ or \mm\ of space cost $o(n)$ and help cost 
$o(n^2)$? We conjecture that the answer is no for both problems. 

\item Is it possible to give a semi-streaming scheme for \triangles\ with a ``commutative'' verifier? We conjecture that the answer is no. If true,
this would contrast with the standard (sans prover) streaming model, where it is known that any streaming algorithm that works in the turnstile update model can be simulated by a linear sketch \cite{woodruffetal}.

\end{itemize}

\medskip

\section*{Acknowledgments} 
The author is grateful to Graham Cormode, Amit Chakrabarti, Andrew McGregor, and Suresh Venkatasubramanian for many valuable discussions regarding this work.

\bibliographystyle{alpha}
\bibliography{bibs}

\newcommand{\etalchar}[1]{$^{#1}$}
\begin{thebibliography}{PTTW13}

\bibitem[AB09]{arorabarak}
Sanjeev Arora and Boaz Barak.
\newblock {\em Computational Complexity: A Modern Approach}.
\newblock Cambridge University Press, New York, NY, USA, 1st edition, 2009.

\bibitem[AGM12a]{graphsketch1}
Kook~Jin Ahn, Sudipto Guha, and Andrew McGregor.
\newblock Analyzing graph structure via linear measurements.
\newblock In Yuval Rabani, editor, {\em SODA}, pages 459--467. SIAM, 2012.

\bibitem[AGM12b]{graphsketch2}
Kook~Jin Ahn, Sudipto Guha, and Andrew McGregor.
\newblock Graph sketches: sparsification, spanners, and subgraphs.
\newblock In Michael Benedikt, Markus Kr{\"o}tzsch, and Maurizio Lenzerini,
  editors, {\em PODS}, pages 5--14. ACM, 2012.

\bibitem[AW09]{aw}
Scott Aaronson and Avi Wigderson.
\newblock Algebrization: A new barrier in complexity theory.
\newblock {\em ACM Trans. Comput. Theory}, 1(1):2:1--2:54, February 2009.

\bibitem[AYZ97]{AlonYZ97}
Noga Alon, Raphael Yuster, and Uri Zwick.
\newblock Finding and counting given length cycles.
\newblock {\em Algorithmica}, 17(3):209--223, 1997.

\bibitem[Bab85]{babai}
L{\'a}szl{\'o} Babai.
\newblock Trading group theory for randomness.
\newblock In Robert Sedgewick, editor, {\em STOC}, pages 421--429. ACM, 1985.

\bibitem[BYKS02]{Bar-YossefKS02}
Ziv Bar-Yossef, Ravi Kumar, and D.~Sivakumar.
\newblock Reductions in streaming algorithms, with an application to counting
  triangles in graphs.
\newblock In David Eppstein, editor, {\em SODA}, pages 623--632. ACM/SIAM,
  2002.

\bibitem[CCGT14]{soda}
Amit Chakrabarti, Graham Cormode, Navin Goyal, and Justin Thaler.
\newblock Annotations for sparse data streams.
\newblock In Chandra Chekuri, editor, {\em SODA}, pages 687--706. SIAM, 2014.

\bibitem[CCM{\etalchar{+}}13]{ccc}
Amit Chakrabarti, Graham Cormode, Andrew McGregor, Justin Thaler, and Suresh
  Venkatasubramanian.
\newblock On interactivity in arthur-merlin communication and stream
  computation.
\newblock {\em Electronic Colloquium on Computational Complexity (ECCC)},
  20:180, 2013.

\bibitem[CCMT14]{icalp}
Amit Chakrabarti, Graham Cormode, Andrew Mcgregor, and Justin Thaler.
\newblock Annotations in data streams.
\newblock {\em Preliminary Version in ICALP 2009. Journal Version to Appear in
  ACM Transactions on Algorithms}, 2014.

\bibitem[CF14]{grahamsampling}
Graham Cormode and Donatella Firmani.
\newblock A unifying framework for ; 0-sampling algorithms.
\newblock {\em Distributed and Parallel Databases}, 32(3):315--335, 2014.

\bibitem[CKLR11]{chung}
Kai-Min Chung, Yael~Tauman Kalai, Feng-Hao Liu, and Ran Raz.
\newblock Memory delegation.
\newblock In Phillip Rogaway, editor, {\em CRYPTO}, volume 6841 of {\em Lecture
  Notes in Computer Science}, pages 151--168. Springer, 2011.

\bibitem[CMT12]{itcs}
Graham Cormode, Michael Mitzenmacher, and Justin Thaler.
\newblock Practical verified computation with streaming interactive proofs.
\newblock In Shafi Goldwasser, editor, {\em ITCS}, pages 90--112. ACM, 2012.

\bibitem[CMT13]{esa}
Graham Cormode, Michael Mitzenmacher, and Justin Thaler.
\newblock Streaming graph computations with a helpful advisor.
\newblock {\em Algorithmica}, 65(2):409--442, 2013.

\bibitem[CTY11]{vldb}
Graham Cormode, Justin Thaler, and Ke~Yi.
\newblock Verifying computations with streaming interactive proofs.
\newblock {\em PVLDB}, 5(1):25--36, 2011.

\bibitem[GKP12]{graphsketch3}
Ashish Goel, Michael Kapralov, and Ian Post.
\newblock Single pass sparsification in the streaming model with edge
  deletions.
\newblock {\em CoRR}, abs/1203.4900, 2012.

\bibitem[GKR08]{gkr}
Shafi Goldwasser, Yael~Tauman Kalai, and Guy~N. Rothblum.
\newblock Delegating computation: interactive proofs for muggles.
\newblock In {\em Proceedings of the 40th Annual ACM Symposium on Theory of
  Computing}, STOC '08, pages 113--122, New York, NY, USA, 2008. ACM.

\bibitem[GM11]{iblt}
Michael~T. Goodrich and Michael Mitzenmacher.
\newblock Invertible bloom lookup tables.
\newblock In {\em Allerton}, pages 792--799. IEEE, 2011.

\bibitem[GMR89]{gmr}
Shafi Goldwasser, Silvio Micali, and Charles Rackoff.
\newblock The knowledge complexity of interactive proof systems.
\newblock {\em SIAM J. Comput.}, 18(1):186--208, 1989.

\bibitem[GR13]{gur}
Tom Gur and Ran Raz.
\newblock Arthur-{M}erlin streaming complexity.
\newblock In {\em Proceedings of the 40th International Colloquium on Automata,
  Languages and Programming: Part I}, ICALP '13, Berlin, Heidelberg, 2013.
  Springer-Verlag.

\bibitem[KP13]{prakash}
Hartmut Klauck and Ved Prakash.
\newblock Streaming computations with a loquacious prover.
\newblock In Robert~D. Kleinberg, editor, {\em ITCS}, pages 305--320. ACM,
  2013.

\bibitem[KP14]{prakashnew}
Hartmut Klauck and Ved Prakash.
\newblock An improved interactive streaming algorithm for the distinct elements
  problem.
\newblock In Javier Esparza, Pierre Fraigniaud, Thore Husfeldt, and Elias
  Koutsoupias, editors, {\em ICALP (1)}, volume 8572 of {\em Lecture Notes in
  Computer Science}, pages 919--930. Springer, 2014.

\bibitem[LFKN92]{lfkn}
Carsten Lund, Lance Fortnow, Howard Karloff, and Noam Nisan.
\newblock Algebraic methods for interactive proof systems.
\newblock {\em J. ACM}, 39:859--868, October 1992.

\bibitem[LMS13]{LeeMS13}
Troy Lee, Fr{\'e}d{\'e}ric Magniez, and Miklos Santha.
\newblock Improved quantum query algorithms for triangle finding and
  associativity testing.
\newblock In Sanjeev Khanna, editor, {\em SODA}, pages 1486--1502. SIAM, 2013.

\bibitem[LNW14]{woodruffetal}
Yi~Li, Huy~L. Nguyen, and David Woodruff.
\newblock Turnstile streaming algorithms might as well be linear sketches.
\newblock In {\em STOC}, 2014.

\bibitem[McG09]{andrew}
Andrew McGregor.
\newblock Graph mining on streams.
\newblock In Ling Liu and M.~Tamer {\"O}zsu, editors, {\em Encyclopedia of
  Database Systems}, pages 1271--1275. Springer, 2009.

\bibitem[McG14]{newsurvey}
Andrew McGregor.
\newblock Graph stream algorithms: A survey.
\newblock {\em SIGMOD Rec.}, 43(1):9--20, May 2014.

\bibitem[Mut05]{muthu}
S.~Muthukrishnan.
\newblock {\em Data Streams: Algorithms And Applications}.
\newblock Foundations and Trends in Theoretical Computer Science. Now
  Publishers Incorporated, 2005.

\bibitem[ope]{openproblem}
List of open problems in sublinear algorithms: Problem 47.
\newblock \url{http://sublinear.info/47}.

\bibitem[PSTY13]{eurocryptdie}
Charalampos Papamanthou, Elaine Shi, Roberto Tamassia, and Ke~Yi.
\newblock Streaming authenticated data structures.
\newblock In Thomas Johansson and Phong~Q. Nguyen, editors, {\em EUROCRYPT},
  volume 7881 of {\em Lecture Notes in Computer Science}, pages 353--370.
  Springer, 2013.

\bibitem[PTTW13]{PavanTTW13}
A.~Pavan, Kanat Tangwongsan, Srikanta Tirthapura, and Kun-Lung Wu.
\newblock Counting and sampling triangles from a graph stream.
\newblock {\em Proc. VLDB Endow.}, 6(14):1870--1881, September 2013.

\bibitem[Sha92]{shamir92}
Adi Shamir.
\newblock {IP = PSPACE}.
\newblock {\em J. ACM}, 39:869--877, October 1992.

\bibitem[SS12]{schroder}
Dominique Schr{\"o}der and Heike Schr{\"o}der.
\newblock Verifiable data streaming.
\newblock In Ting Yu, George Danezis, and Virgil~D. Gligor, editors, {\em ACM
  Conference on Computer and Communications Security}, pages 953--964. ACM,
  2012.

\bibitem[SV11]{SuriV11}
Siddharth Suri and Sergei Vassilvitskii.
\newblock Counting triangles and the curse of the last reducer.
\newblock In Sadagopan Srinivasan, Krithi Ramamritham, Arun Kumar, M.~P.
  Ravindra, Elisa Bertino, and Ravi Kumar, editors, {\em WWW}, pages 607--614.
  ACM, 2011.

\bibitem[Tha13]{thaler}
Justin Thaler.
\newblock Time-optimal interactive proofs for circuit evaluation.
\newblock In Ran Canetti and Juan~A. Garay, editors, {\em CRYPTO (2)}, volume
  8043 of {\em Lecture Notes in Computer Science}, pages 71--89. Springer,
  2013.

\bibitem[VSBW13]{allspice}
Victor Vu, Srinath Setty, Andrew~J. Blumberg, and Michael Walfish.
\newblock A hybrid architecture for interactive verifiable computation.
\newblock In {\em IEEE Symposium on Security and Privacy}, 2013.

\end{thebibliography}

\appendix
\section{Extensions: Counting Structures Other Than Triangles}
\label{app:extend}
Let $H$ be a graph on $k$ vertices. It is possible to extend the protocol underlying Theorem \ref{thm:triangles} to count the number of occurrences of $H$ as a subgraph of $G$. The protocol requires $k - 2$ rounds, and its help and space costs are $O(k^3 n \log n)$ and $O(kn\log n)$ respectively. For concreteness, we describe the protocol in detail for the case where $H$ is a 4-cycle.

\medskip \noindent \textbf{Protocol for Counting 4-Cycles.}
As in the proof of Theorem \ref{thm:triangles}, let $G_i$ denote the graph defined by the first $i$ stream updates $\langle (e_1 , \Delta_1 ), \dots , (e_i , \Delta_i )\rangle$, and let $E_i \colon [n] \times [n] \rightarrow \mathbb{Z}$ denote the function that outputs the multiplicity of the edge $(u,v)$ in graph $G_{i-1}$. On edge update 
$e_i = (u_i, v_i)$, notice that the number of four-cycles that $e_i$ completes in $E_i$ is precisely 
$\Delta_i \cdot \sum_{z_1,z_2 \in [n]} E_i(u_i,z_1)E_i(z_1,z_2)E_i(z_2,v_i).$ (Technically, this equality holds only assuming that there are no self-loops in $G$. However, the protocol is easily modified to handle graphs $G$ (and sub-graphs $H$) that both may have self-loops). Thus, at the end of the stream, the total number of
4-cycles in the graph $G = G_m$ is precisely
\begin{align} \notag 
\sum_{i\leq m} \Delta_i \sum_{z_1, z_2 \in [n]} E_i(u_i,z_1)E_i(z_1,z_2)E_i(z_2,v_i) =\\
\notag \sum_{z_1, z_2 \in [n]} \sum_{i\leq m}  \Delta_i E_i(u_i,z_1)E_i(z_1,z_2)E_i(z_2,v_i)  = \\
\sum_{z_1, z_2 \in [n]} \sum_{i\leq m}  \Delta_i \tilde{E}_i(u_i,z_1)\tilde{E}_i(z_1,z_2)\tilde{E}_i(z_2,v_i).
\label{screwdropbox}
\end{align}
In Expression \eqref{screwdropbox}, the polynomial $\tilde{E}_i$ is defined exactly as in the proof of Theorem \ref{thm:triangles}, except the field $\mathbb{F}$ over which $\tilde{E}_i$ is defined must have size $12(Bn)^4 \leq |\mathbb{F}| ² 24(Bn)^4$ (the reason for the larger field is simply to ensure that the field is large enough to represent the maximum possible number of 4-cycles in the graph, without ``wrap-around'' issues). 
To compute Expression \eqref{screwdropbox}, the prover and verifier run the sum-check protocol of Lund et al. \cite{lfkn} --- specifically, they apply the sum-check protocol to the bivariate polynomial
$$g(Z_1,Z_2) := \sum_{i \leq m} \Delta_i \tilde{E}_i(u_i, Z_1) \tilde{E}_i(Z_1,Z_2)\tilde{E}_i(Z_2,v_i).$$
%We direct the interested reader to \cite[Chapter 8]{arorabarak} for a detailed description of the sum-check protocol. For our purposes, the crucial property of the sum-check protocol is the following: in order to run her part of the protocol, the verifier randomly chooses two values $r_1, r_2$ at random from $\mathbb{F}$, and needs to evaluate the quantity 
 %$g(r_1, r_2) = \sum_{i \leq m} \Delta_i \tilde{E}_i(u_i, r_1) \tilde{E}_i(r_1,r_2)\tilde{E}_i(r_2,v_i)$. 
Using techniques identical to those used in Theorem \ref{thm:triangles}, one can show that,  as required by the sum-check protocol, the verifier can compute $g(r_1, r_2)$ in space $O(n \log |\mathbb{F}|)$ with a single streaming pass over the input.

\medskip
\noindent \textbf{Summary of Costs.}	 Recall that the sum-check protocol applied a $k$-variate polynomial $g(Z_1, \dots, Z_k)$ requires $2k-1$ messages (more precisely, it requires the prover to send a total of $k$ messages to the verifier, and the verifier to send a total of $k - 1$ responses to the prover). The verifier's $i$th message to the prover consists of the single field element $r_i \in \mathbb{F}$, while the prover's $i$'th message to the verifier is univariate polynomial $s_i(Z_i)$ whose degree is at most $\deg_i(g)$, the degree of $g$ in the variable $Z_i$.

In our context, this translates to a 3-message protocol for counting 4-cycles, in which the total communication cost is $O(n \log n)$. The soundness parameter $\delta_s$ of the protocol is at most $\sum_i \deg_i(g)/|\mathbb{F}| \leq 4n/|\mathbb{F}| \leq B^4/(3n^3)$.

More generally, given any $k$-node subgraph $H$, we can count the number of occurrences $H$ in $G$ by using a suitable $(k - 2)$-variate polynomial $g$. Applying the sum-check protocol to $g$ requires $2k-1$ messages (spread over $k$ rounds) between the prover and verifier. The space usage of the verifier to evaluate g at a random point becomes $O(k n \log n)$, and the total communication cost is $O(k^3 n \log n)$. Here, the factor of $k$ in the verifier's space usage is due to the need to work over a field whose size grows exponentially in $k$. The factor of $k^3$ in the communication complexity is due to the following three sources: one factor of $k$ is due to the need to work in a field whose size grows exponentially with $k$, another factor is due to the fact that $\deg_i(g)$ grows linearly with $k$, for each variable $Z_i$, and the final factor of $k$ is due to the fact that the number of messages sent by the prover grows linearly with $k$.

\end{document}